\documentclass[preprint,preprintnumbers,amssymb,latexsym,nofootinbib,array,enumerate,letter,superscriptaddress,11pt]{revtex4}

\pdfoutput=1
\usepackage{graphicx}
\usepackage{epstopdf}
\usepackage{dcolumn}
\usepackage{bm}
\usepackage{hyperref}
\usepackage{amsmath}

\begin{document}


\def\a{\alpha}
\def\b{\beta}
\def\c{\varepsilon}
\def\d{\delta}
\def\e{\epsilon}
\def\f{\phi}
\def\g{\gamma}
\def\h{\theta}
\def\k{\kappa}
\def\l{\lambda}
\def\m{\mu}
\def\n{\nu}
\def\p{\psi}
\def\q{\partial}
\def\r{\rho}
\def\s{\sigma}
\def\t{\tau}
\def\u{\upsilon}
\def\v{\varphi}
\def\w{\omega}
\def\x{\xi}
\def\y{\eta}
\def\z{\zeta}
\def\D{\Delta}
\def\G{\Gamma}
\def\H{\Theta}
\def\L{\Lambda}
\def\F{\Phi}
\def\P{\Psi}
\def\S{\Sigma}

\def\o{\over}
\def\beq{\begin{align}}
\def\eeq{\end{align}}
\newcommand{\gsim}{ \mathop{}_{\textstyle \sim}^{\textstyle >} }
\newcommand{\lsim}{ \mathop{}_{\textstyle \sim}^{\textstyle <} }
\newcommand{\vev}[1]{ \left\langle {#1} \right\rangle }
\newcommand{\bra}[1]{ \langle {#1} | }
\newcommand{\ket}[1]{ | {#1} \rangle }
\newcommand{\mEV}{ {\rm meV} }
\newcommand{\EV}{ {\rm eV} }
\newcommand{\KEV}{ {\rm keV} }
\newcommand{\MEV}{ {\rm MeV} }
\newcommand{\GEV}{ {\rm GeV} }
\newcommand{\TEV}{ {\rm TeV} }
\newcommand{\1}{\mbox{1}\hspace{-0.25em}\mbox{l}}
\newcommand{\headline}[1]{\noindent{\bf #1}}
\def\diag{\mathop{\rm diag}\nolimits}
\def\Spin{\mathop{\rm Spin}}
\def\SO{\mathop{\rm SO}}
\def\O{\mathop{\rm O}}
\def\SU{\mathop{\rm SU}}
\def\U{\mathop{\rm U}}
\def\Sp{\mathop{\rm Sp}}
\def\SL{\mathop{\rm SL}}
\def\tr{\mathop{\rm tr}}
\def\mpl{M_{\rm Pl}}
\renewcommand{\arraystretch}{1.5}

\def\IJMP{Int.~J.~Mod.~Phys. }
\def\MPL{Mod.~Phys.~Lett. }
\def\NP{Nucl.~Phys. }
\def\PL{Phys.~Lett. }

\def\PR{Phys.~Rev. }
\def\PRL{Phys.~Rev.~Lett. }
\def\PTP{Prog.~Theor.~Phys. }
\def\ZP{Z.~Phys. }

\def\dd{\mathrm{d}}
\def\ff{\mathrm{f}}
\def\BH{{\rm BH}}
\def\inf{{\rm inf}}
\def\ev{{\rm evap}}
\def\eq{{\rm eq}}
\def\SM{{\rm sm}}
\def\Mpl{M_{\rm Pl}}
\def\GeV{{\rm GeV}}

\def\atp#1{{ \bf \textcolor{blue}{[AP: {#1}]} }}
\def\zj#1{{\textcolor{Plum}{[ZJ: {#1}]} }}
\def\comment#1{{\textcolor{Red}{[Comment: {#1}]} }}

\def\newpar{\vskip4pt}

\title{Simple Hidden Sector Dark Matter}
\preprint{LCTP-20-05}
\preprint{CERN-TH-2020-024}

\author{Patrick Barnes}
\affiliation{Leinweber Center for Theoretical Physics, Department of Physics, University of Michigan, Ann Arbor, MI 48109, USA}
\author{Zachary Johnson}
\affiliation{Leinweber Center for Theoretical Physics, Department of Physics, University of Michigan, Ann Arbor, MI 48109, USA}
\author{Aaron Pierce}
\affiliation{Leinweber Center for Theoretical Physics, Department of Physics, University of Michigan, Ann Arbor, MI 48109, USA}
\author{Bibhushan Shakya}
\affiliation{CERN, Theoretical Physics Department, 1211 Geneva 23, Switzerland}

\begin{abstract}
A hidden sector that kinetically mixes with the Minimal Supersymmetric Standard Model provides simple and well-motivated dark matter candidates that possess many of the properties of a traditional weakly interacting massive particle (WIMP). These supersymmetric constructions can also provide a natural explanation for why the dark matter is at the weak scale – even if it resides in a hidden sector. In the hidden sector, a natural pattern of symmetry breaking generally makes particles and their superpartners lie around the same mass scale, opening novel possibilities for a variety of cosmological histories and complex indirect detection signatures.
\end{abstract}

\maketitle

\tableofcontents

\noindent

\section{Motivation}\label{sec:motivation}

A thermally produced stable particle with a weak scale annihilation cross section reproduces the observed dark matter relic abundance \cite{Lee:1977ua}. The weak scale is known to be an important scale from the particle physics perspective, and many beyond the Standard Model (BSM) theories have dark matter (DM) candidates at this scale, often as part of a solution to the hierarchy problem.  This coincidence has been dubbed the weakly interacting massive particle (WIMP) miracle, and thermal DM associated with hierarchy problem solutions has been the subject of intense theoretical research as well as experimental searches.  An examplar is the lightest supersymmetric particle (LSP) \cite{Goldberg:1983nd}, stabilized by R-parity.  However, even though several compelling arguments --  ranging from gauge coupling unification to considerations of the underlying theory of quantum gravity --  provide reasons to believe that supersymmetry is part of the underlying description of nature, the absence of  signals at DM direct detection experiments such as Xenon1T \cite{Aprile:2018dbl} and the non-observation of superpartners at the Large Hadron Collider (LHC) have led to questions about whether the simplest instantiation of this WIMP DM idea is realized in nature \cite{Perelstein:2011tg,Amsel:2011at,Perelstein:2012qg}.

While it is possible that a weak scale cross section is associated with the Standard Model (SM) weak interactions themselves \cite{Cirelli:2009uv, Cohen:2011ec}, it is not necessary.  New dynamics associated with the DM particle may be unrelated to the SM and hold no direct connection to a  hierarchy problem solution.  In particular, the WIMP miracle can be realized with order one  couplings within a separate weak scale dark \emph{sector} that only interacts very feebly with the SM.  This idea was dubbed secluded dark matter in \cite{Pospelov:2007mp}. The existence of such secluded/hidden/dark sectors, with extended gauge groups, is well-motivated from a string theory perspective, and their interactions with the SM can give rise to several interesting phenomenological signatures (see \cite{Alexander:2016aln} and references therein). In particular, the kinetic mixing portal \cite{Holdom:1985ag}, where a gauged $U(1)'$ in a hidden sector mixes with the SM hypercharge $U(1)_Y$ \cite{Langacker:2008yv}, has been extensively studied in the literature, including in the context of dark matter that realizes its thermal abundance via the aforementioned ``WIMP" miracle, see, e.g., \cite{Pospelov:2007mp,Feldman:2006wd,Bell:2016fqf,Bell:2016uhg,Evans:2017kti}.

If the DM is near the weak scale but in a separate sector, it is of interest to understand how that sector knows about the weak scale.  In supersymmetric theories, this may occur naturally if supersymmetry breaking is mediated to both sectors with approximately equal strength, as might happen, e.g., in theories of gravity mediation.  In this case, the masses in the two sectors are correlated at some UV scale but may be separated at lower energies by running effects.   We use this line of argument to motivate hidden sector spectra. While hidden sector particles such as heavy $Z^{\prime}$'s are difficult to probe directly,\footnote{For studies of possible phenomenological implications of hidden sector gauge bosons and their superpartners, see \cite{Arvanitaki:2009hb,Baryakhtar:2012rz}).} the existence of such a  (largely) hidden sector could have consequences for cosmology, in particular  for dark matter.  Here we will work under the assumption that the two sectors are coupled strongly enough that the hidden and visible sectors thermalize.

While this broad-brush picture has some appeal, it is of interest to ask whether the data from the LHC can tell us more about such a supersymmetric setup. The absence of superpartners at the LHC suggests  the weak scale itself is somewhat fine-tuned, as
does the Higgs boson mass, which requires large loop level corrections due to supersymmetry breaking \cite{Haber:1990aw,Ellis:1990nz,Okada:1990vk}.
 In fact, the relatively large value of the observed Higgs boson mass suggests that the scale of supersymmetry breaking is several TeV in the absence of significant stop mixing in the Minimal Supersymmetric Standard Model (MSSM). This ``little hierarchy problem" -- wherein the  weak scale is tuned at a sub-percent level --  might simply be accidental, or find explanations in some anthropic or cosmological selection process.  None of these need apply to the hidden sector, and as a consequence the vacuum expectation value (vev) in the hidden sector should be more closely tied to the scale of supersymmetry breaking. 

WIMP dark matter in minimal supersymmetric setups is made stable by assuming R-parity. This, however, does not work for a hidden sector dark matter candidate if the hidden sector spectrum is heavier than that of the visible sector, since the said dark matter candidate can decay into the visible sector even in the presence of R-parity. In this case, DM can instead be stabilized by other, perhaps accidental, symmetries realized in the hidden sector.  
Interestingly, R-parity need not be conserved, and the breaking of R-parity might even be desirable, for instance, to break baryon number in order to realize baryogenesis, as studied in \cite{Pierce:2019ozl}. 

In this paper, we combine the above ideas to construct simple and realistic  models for hidden sector dark matter. We assume this sector interacts with our own via supersymmetric kinetic mixing.  All the ingredients -- hidden sectors, supersymmetry, and kinetic mixing -- are well-motivated. 
In the absence of accidental tuning in the hidden sector, no large mass hierarchies are expected between hidden sector particles and their superpartners, so that a multitude of particles can be involved in both dark matter freeze-out as well as present day dark matter annihilation.\footnote{For related studies of dark matter in very supersymmetric hidden sectors, see \cite{Dery:2019jwf}.} Our study therefore illuminates the wide range of dynamics that can give rise to a WIMP-like miracle in well-motivated hidden sectors.   

The outline of the paper is as follows: In Section\,\ref{sec:framework}, we describe the field content of the hidden sector we consider, outlining the possible dark matter candidates. This is followed by detailed studies of  fermion and scalar dark matter in Sections\,\ref{sec:fermion} and \ref{sec:scalar}, respectively. Section\,\,\ref{sec:coannihilations} addresses cases where the mass gap between the dark matter candidate and its superpartner is small, leading to coannihilation effects and long lifetimes. We then explore relations between parameters in the UV and IR in Section\,\ref{sec:UV}, discussing how consistent cosmological histories can emerge from reasonable parameter choices in the UV. Section \ref{sec:decays} is devoted to the discussion of the decay modes of various hidden sector particles. Direct detection and collider constraints are explored in Section\,\ref{sec:direct}, followed by a discussion of indirect detection signals in Section\,\ref{sec:indirect}. We end with some concluding remarks in Section\,\ref{sec:conclusion}.

\section{A Simple Dark Sector}\label{sec:framework}

In addition to the field content of the MSSM, we consider a dark/hidden sector with gauge group $U(1)'$ and a trio of SM-singlet superfields: a dark Higgs field $\hat{H}^{\prime}$ with charge  $Q^{\prime} = +1$ (which breaks the $U(1)'$ symmetry once the scalar component obtains a vev), a superfield $\hat{T}$ with charge $Q^{\prime} = -1$ (necessary for cancellation of anomalies related to $U(1)'$), and a singlet superfield $\hat{S}$ with charge $Q^{\prime} = 0$ (necessary to enable a scale-invariant superpotential involving $\hat{H}^{\prime}$, $\hat{T}^{\prime}$). 
The most general superpotential after imposing the above $U(1)'$ charges along with any symmetry under which both $\hat{S}$ and $\hat{T}$ transform non-trivially is
\begin{equation}
{\mathcal W}_{hid} = \lambda \hat{S} \hat{T} \hat{H}^{\prime}.
\end{equation} 
This superpotential possesses a $\mathbb{Z}_2$ symmetry under which both $\hat{S}$ and $\hat{T}$ are odd; this ensures the lightest particle in the $\hat{S}-\hat{T}$ system, which we will refer to as the lightest $\mathbb{Z}_{2}$ odd particle (LZP), is stable and therefore a dark matter candidate.\footnote{In contrast, a pure singlet superfield $\hat{S}$ that can couple to SM fields would give rise to decaying dark matter via a neutrino portal, see e.g. \cite{Roland:2014vba,Roland:2015yoa,Shakya:2015xnx,Roland:2016gli,Shakya:2016oxf}.} We assume that the hidden sector communicates with the visible sector via supersymmetric kinetic mixing \cite{Dienes:1996zr}:
\begin{equation}
\label{eq:Kmixing}
\frac{\epsilon}{2} \int d^{2} \theta \,  W_{Y} W^{\prime} + h.c.  = \epsilon D_{Y} D^{\prime} - \frac{\epsilon}{2} F_Y^{\mu \nu}  F^{\prime}_{\mu \nu} + i \epsilon \tilde{B} \sigma^{\mu} \partial_{\mu} {\tilde{B}}^{\prime \, \dagger}
+ i \epsilon \tilde{B}^{\prime} \sigma^{\mu} \partial_{\mu} \tilde{B}^{\dagger}, 
\end{equation}
where the $W_{Y}$, $W^{\prime}$ represent the chiral field strength multiplet for $U(1)_Y$ hypercharge and the hidden sector $U(1)'$, respectively, and we use the notation $\tilde{B'}$ for the hidden sector gaugino. This basic set-up has previously also been considered in the context of asymmetric dark matter \cite{Cohen:2010kn}, as well as a way to generate dark matter at the GeV scale \cite{Morrissey:2009ur}.  It was also considered in some detail in \cite{Andreas:2011in}, where some consequences for thermal histories and direct detection were considered. 

We assume supersymmetry breaking induces a vev for $H^{\prime}$ only, $\langle H^{\prime} \rangle = v^{\prime}/\sqrt{2}$. This vacuum will be preferred when there  is either a hierarchy between the soft masses for the scalars, or if $\lambda$ is large enough to overcome the D-flatness condition (which favors $\langle H^{\prime} \rangle = \langle T \rangle$).    This vev provides the hidden gauge boson $Z'$ with a mass $m_{Z'} = g^{\prime} v^{\prime}$ and combines the fermion components of $\hat{S}$ and $\hat{T}$ into a Dirac fermion, which we denote $\psi$, with $m_{\psi} =\lambda v^{\prime}/\sqrt{2}$. As we will discuss later (Sec.~\ref{sec:UV}), $\alpha_\lambda=2\alpha'$ is an RG fixed point where an accidental ${\mathcal N}$ = 2 SUSY is restored; at this point the $\psi, Z',$ and $H'$ are degenerate.  

The hidden neutralino sector has the following mass matrix in the $\tilde{B'}$, $\tilde{H^{\prime}}$ basis:
\begin{equation}
\label{eqn:neumat}
{\mathcal M}_{\chi^{\prime}} =
\begin{pmatrix}
m_{\tilde{B'}} & m_{Z'} \\
m_{Z'} & 0 
\end{pmatrix},
\end{equation}
with $m_{\tilde{B'}}$ the hidden sector gaugino mass. In the supersymmetric limit $m_{\tilde{B'}} \rightarrow 0$   (also taking $\epsilon \rightarrow 0$), $\tilde{B'}$ pairs with $\tilde{H^{\prime}}$ to form a Dirac neutralino that is degenerate with $Z^{\prime}$. A nonzero $m_{\tilde{B'}}$ splits this state into two Majorana mass eigenstates, which we denote $\chi_{1}^{\prime}$ and $\chi_{2}^{\prime}$, with $m_{\chi_{1}^{\prime}}<m_{\chi_{2}^{\prime}}$. If $m_{\tilde{B'}}< m_{Z'}$, the mass splitting is small and the mass eigenstates contain significant $\tilde{B'}-\tilde{H}'$ admixtures; on the other hand, the hierarchy $m_{\tilde{B'}}\gg m_{Z^{\prime}}$ represents a seesaw limit where the lightest eigenstate is approximately $\tilde{H}$ with suppressed mass $|m_{\chi_{1}^\prime}| \approx m_{Z'}^2/m_{\tilde{B'}}$.  For later convenience, we define a mixing angle $\theta_N$, with $\chi_{1}^{\prime} = \cos{\theta_N} \tilde{H}^{\prime} - \sin{\theta_{N}} \tilde{B}^{\prime}.$ 

In the extended neutralino sector, we also allow a gaugino mass portal
\begin{equation}
\label{eqn:gauginomassmix}
{\mathcal L} \supset - \epsilon \,m_{\tilde{B} \tilde{B'}} \tilde{B} \tilde{B'} + \rm{h.c.},
\end{equation}
where we have pulled out a factor of $\epsilon$ to emphasize that we expect mass mixing of this order.   
In the Higgs sector, upon elimination of the auxilliary fields, we have a $D$-term contribution to the Higgs potential that includes:
\begin{equation} 
V_D \ni (\frac{g^{2}}{8}+\frac{g_{Y}^2}{8 (1- \epsilon^2)} )( |H_{u}|^2-|H_{d}|^2)^2
+\frac{ g^{\prime \, 2}}{2 (1- \epsilon^2)} |H^{\prime}|^4 -  \frac{\epsilon}{2(1-\epsilon^2)} g^{\prime} g_{Y} (|H_{u}|^2 - |H_{d}|^2) |H^{\prime}|^2.
\label{eq:Dterm}
\end{equation}
Thus, the kinetic mixing also provides a Higgs portal between the two sectors.

In the supersymmetric limit, the hidden Higgs boson $H'$ is degenerate with the $Z^{\prime}$.  Due to supersymmetry breaking effects, the $H'$ mass receives loop corrections analogous to the well-known top loop correction in the MSSM \cite{Haber:1990aw,Ellis:1990nz,Okada:1990vk}.   The size of this correction will depend on the hidden sector couplings $\lambda$ and $g^{\prime}$. We have redone this one-loop calculation to the Higgs mass in the effective potential formalism. In general, we find that the correction is modest, since the logarithm is smaller due to smaller mass splittings between superpartners ($\textit{i.e.}$, because we assume less tuning) in the hidden sector,  and since large values of $\lambda$ in the UV will  rapidly flow to the fixed point $\lambda = \sqrt{2} g^{\prime}$, causing the Yukawa and gauge corrections to the hidden sector Higgs mass to partially cancel.   The corrections can become significant for large values of $\lambda$, particularly in the case $\lambda \gg g^{\prime}$, but we note that this occurs in a region of parameter space where  fine-tuning in the hidden sector is severe. In what follows, we therefore generally assume the tree level relation $m_{H^{\prime}} = m_{Z^{\prime}}$, and comment on places where this assumption may fail.  Finally, the kinetic mixing induces corrections to the mass eigenvalues of both $Z'$ and $H'$, but these are generally quite small.

In the hidden scalar sector, the supersymmetry breaking soft terms are
\begin{equation}
{\mathcal L} \supset \tilde{m}_{S}^2 |\tilde{S}|^2 +\tilde{m}_{T}^2 |\tilde{T}|^2  + \tilde{m}_{H'}^2 |\tilde{H'}|^2+ (\lambda A_{\lambda} \tilde{S} \tilde{T} \tilde{H}' + h.c.).
\end{equation}
In the $(\tilde{S},\ \tilde{T}^*$) basis, the scalar mass matrix can be written as
\begin{equation}
m_{scalar}^2 = \begin{pmatrix}
\tilde{m}_{S}^2  + m_{\psi}^2 & m_{\psi}^* A^*_{\lambda} \\
m_{\psi} A_{\lambda} & \tilde{m}_{T}^2 +m_{\psi}^2 - \frac{1}{2}{m_{Z'}^2}
\end{pmatrix}.
\label{eqn:ScalarMassMatrix}
\end{equation}
We denote the scalar mass eigenstates as $S_1$ and $S_2$, with $m_{S_{1}}< m_{S_{2}}$.  We define a scalar mixing angle $\theta_{S}$ with $S_{1} = \cos{\theta_{S}} \tilde{T}^{\ast} - \sin{\theta_{S}} \tilde{S}$. We follow the convention where both $\lambda$ and $A_\lambda$ are real, but note that the model possesses a  physical phase, $Arg(m_{\tilde{B'}} A_{\lambda}^{\ast})$, which we denote $\phi_{CP}$. Depending on the sizes of the various soft masses, the LZP may be the scalar $S_1$ or fermion $\psi$.

\subsection{Dark Matter Candidates}
Depending on whether R-parity is conserved or broken, several dark matter scenarios are possible. Here, we outline some possibilities before focusing on the R-parity violating (RPV) case for the rest of the paper. For simplicity, we take the LSP to be the visible sector $\tilde{B}$ and assume $m_{\tilde{B}}<m_{\chi_1'}$.\,\footnote{Cosmological aspects of setups with hidden sector gaugino LSP dark matter have been studied in \cite{Ibarra:2008kn,Arvanitaki:2009hb,Acharya:2016fge}.} We assume that the gravitino is sufficiently heavy that it does not affect cosmology. 

If R-parity is conserved, the LSP is stable and therefore another dark matter component in addition to the LZP. If the LSP is lighter than the LZP but freezes out 
before the LZP,  LZP annihilations produce a secondary population of LSP DM. While this provides a contribution to the abundance of LSP DM on top of the thermal abundance, the LSP dark matter population typically retains a thermal distribution since it maintains kinetic equilibrium with the SM bath at the time of LZP decoupling. 
An interesting wrinkle occurs if $\chi_1'$ is sufficiently long-lived (due to small $\epsilon$). 
In this case, $\chi_1'$ can freeze-out prior to the LSP, but decay after LSP freeze-out, contributing another secondary LSP DM abundance (since each $\chi_1'$ decay produces an LSP).   In principle, the DM from $\chi_1'$ decay might give a too-large DM abundance. However, such concerns are mitigated because the $\chi_1'-\tilde{B}$ coannihilation process, which can determine $\chi_1'$ freeze-out, while $\epsilon$ suppressed, can remain in equilibrium  longer than naively expected because of the relatively unsuppressed $\tilde{B}$ abundance. The result is that it is not difficult to suppress the $\chi_1'$ freeze-out abundance (and hence the secondary LSP abundance) to acceptable levels.  

Another interesting possibility is the existence of a trio of dark matter states. If the mass splitting between the LZP and its superpartner is smaller than the LSP mass, then decays between the two are kinematically forbidden.  In this case, the LZP, its superpartner, and the LSP are all stable components of dark matter. 

On the other hand, if R-parity is broken, the LSP decays into SM particles via RPV interactions.  For concreteness, consider the baryon number violating coupling:
\begin{equation}
W_{RPV} = \lambda{''}_{ijk} U_{i}^{c} D_{j}^{c} D_{k}^{c}.
\end{equation}
If $\lambda{''}$ is small, the consequently long lifetime of the LSP is a potential concern, since the LSP abundance can grow to dominate the energy density of the Universe, and the significant entropy from its subsequent decays may dilute the abundance of LZP dark matter. 
While viable cosmologies of this type may be constructed, significant dilution would spoil the ``WIMP miracle" that this scenario realizes.  

In the remainder of this paper, we only consider scenarios where R-parity is broken. Thus both the LZP superpartner and the LSP are unstable, and the LZP is the sole DM candidate. In the next two sections, we discuss a variety of possible spectra with fermion and scalar LZP dark matter, respectively, addressing cosmological histories and present day annihilation cross sections. 

\section{Fermion Dark Matter}\label{sec:fermion}
We first review the cosmology of fermion LZP $\psi$ DM freeze-out with simplified analytic expressions to understand the broad picture, followed by detailed numerical treatment to include more complicated cases. 
 
$\psi$ LZPs can annihilate via $s$-wave processes within the dark sector unless $\psi$ is the lightest dark sector state. 
Over much of the parameter space, the $Z'H'$ channel dominates if open ($\alpha_\lambda>2\alpha'$); recall that $\alpha_\lambda=2\alpha'$ is an RG fixed point.  The $H'H'$ channel is $p$-wave suppressed, and the $Z^{\prime}Z^{\prime}$ channel is suppressed by $\a' / \a_\l$ relative to $Z'H'$.\footnote{For the non-supersymmetric case the importance of the $Z^{\prime} H^{\prime}$ channel was discussed in \cite{Bell:2016fqf,Bell:2016uhg}, assuming both the $Z'$ and $H'$ receive their mass from the Higgs mechanism.}
When $Z^{\prime}H^{\prime}$ is kinematically forbidden, the only other channel potentially available completely within the dark sector is $\chi_1'\chi_1'$.   Annihilations to $Z^{\prime}H^{\prime}$ proceed either via $s$-channel $Z^{\prime}$ exchange or $t$/$u$-channel $\psi$ exchange, while annihilations to $\chi_1^{\prime} \chi_1^{\prime}$ proceed either via  $s$-channel $Z^{\prime}$ or $t$/$u$-channel scalar exchange.
In the limit where the scalars are decoupled, the annihilation cross sections are
\begin{align}
	\langle \sigma v \rangle_{\chi_1' \chi_1'} &\approx c_{\theta_N}^4 \frac{\pi \a'^2  }{ m_\psi^2} \sqrt{1-\eta_{\chi_1'}}\frac{16 \eta_{\chi_1'} + 2 \eta_{Z^{\prime}}^2 - \eta_{\chi_1}\eta_{Z^{\prime}}(8 + \eta_{Z^{\prime}} ) }{\eta_{Z^{\prime}}^2  (\eta_{Z^{\prime}} -4)^2}\,,\\
	\langle \sigma v \rangle_{Z^{\prime} H^{\prime}} &\approx \frac{\pi \a_\l^2  }{4m_\psi^2} \frac{(1-\eta_{Z^{\prime}})^{1/2}( 64 -128\eta_{Z^{\prime}} +104\eta_{Z^{\prime}}^2 -30\eta_{Z^{\prime}}^3 +\eta_{Z^{\prime}}^4+\eta_{Z^{\prime}}^5) }{ (2-\eta_{Z^{\prime}})^2(4-\eta_{Z^{\prime}})^2},
\end{align}
where $\eta_{Z^{\prime}}\equiv m_{Z'}^2/m_\psi^2= 2\alpha'/\alpha_\lambda$, so that $\eta_{Z^{\prime}}=1$ represents the IR fixed point;  $\eta_{\chi_1'}\equiv m_{\chi_1'}^2/m_\psi^2$, and $\theta_{N}$ is the neutralino mixing angle as defined below Eq.~(\ref{eqn:neumat}). 

Using the above expressions, we can compute the approximate dark matter abundance in the specific cases where individual annihilation channels dominate the freeze-out process: 
\begin{align}
    (\Omega h^2)_{\chi_1' \chi_1'} &\approx 
        \Omega_{DM}h^2  \left(\frac{m_\psi}{1 \text{TeV}}\right)^2 \left(\frac{0.05}{\a'}\right)^2 
        \left(\frac{  \eta_{Z^{\prime}} - 4}{  2}\right)^2~, \\ 
        (\Omega h^2)_{Z'H'} &\approx\Omega_{DM}h^2  \left(\frac{m_\psi}{1 \text{TeV}}\right)^2 \left(\frac{0.07}{\a_\l}\right)^2 ~,
\end{align}
where $\Omega_{DM}h^2$ represents the experimentally observed value.
We can use these to infer  approximate combinations of masses and couplings that reproduce the observed dark matter relic density. Excepting the case where the intermediate $Z^{\prime}$ is nearly on resonance, requiring  perturbativity up to the GUT scale imposes the bound $m_\psi \lesssim $ 2-3 TeV. 
Somewhat higher masses are possible in regions of parameter space where multiple channels contribute to dark matter annihilation.

\begin{figure}[t]
	\centering
	\begin{minipage}{.5\textwidth}
	\includegraphics[width=1\columnwidth]{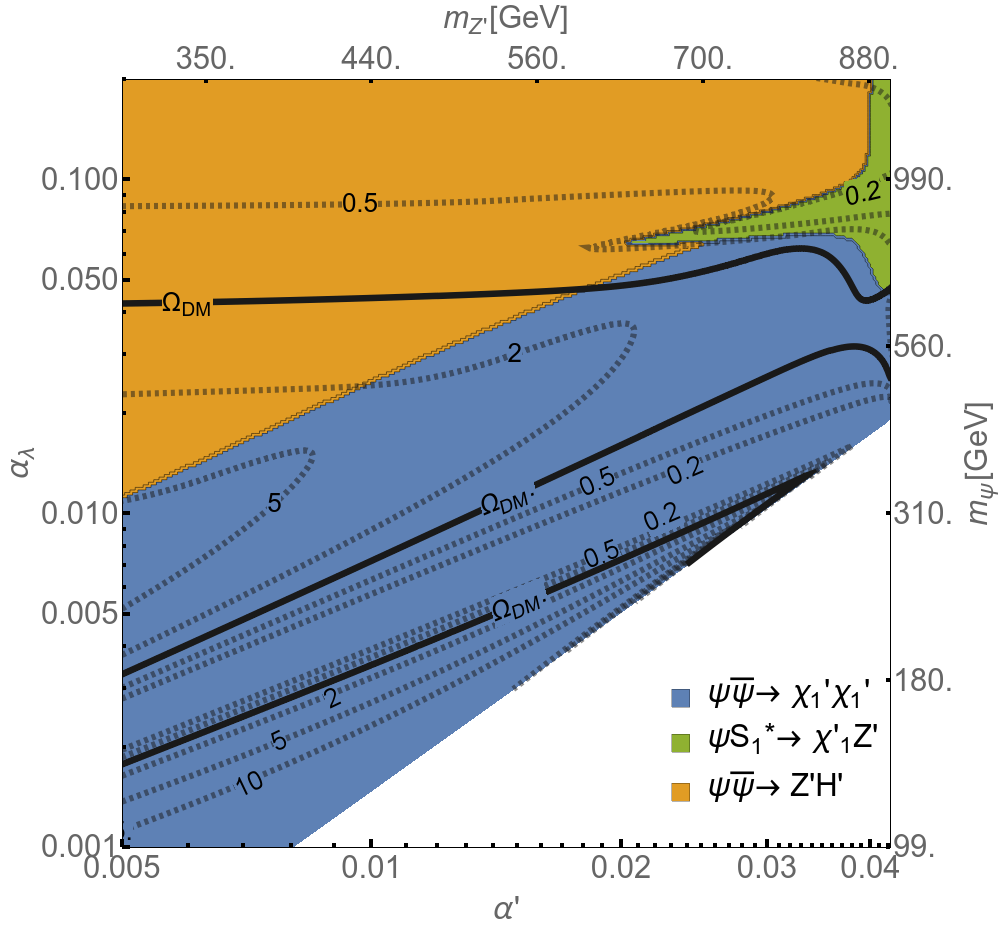}\\\textbf{a) $\Omega/\Omega_{DM}$}
	\end{minipage}%
	~~
	\begin{minipage}{.5\textwidth}
	\includegraphics[width=1\columnwidth]{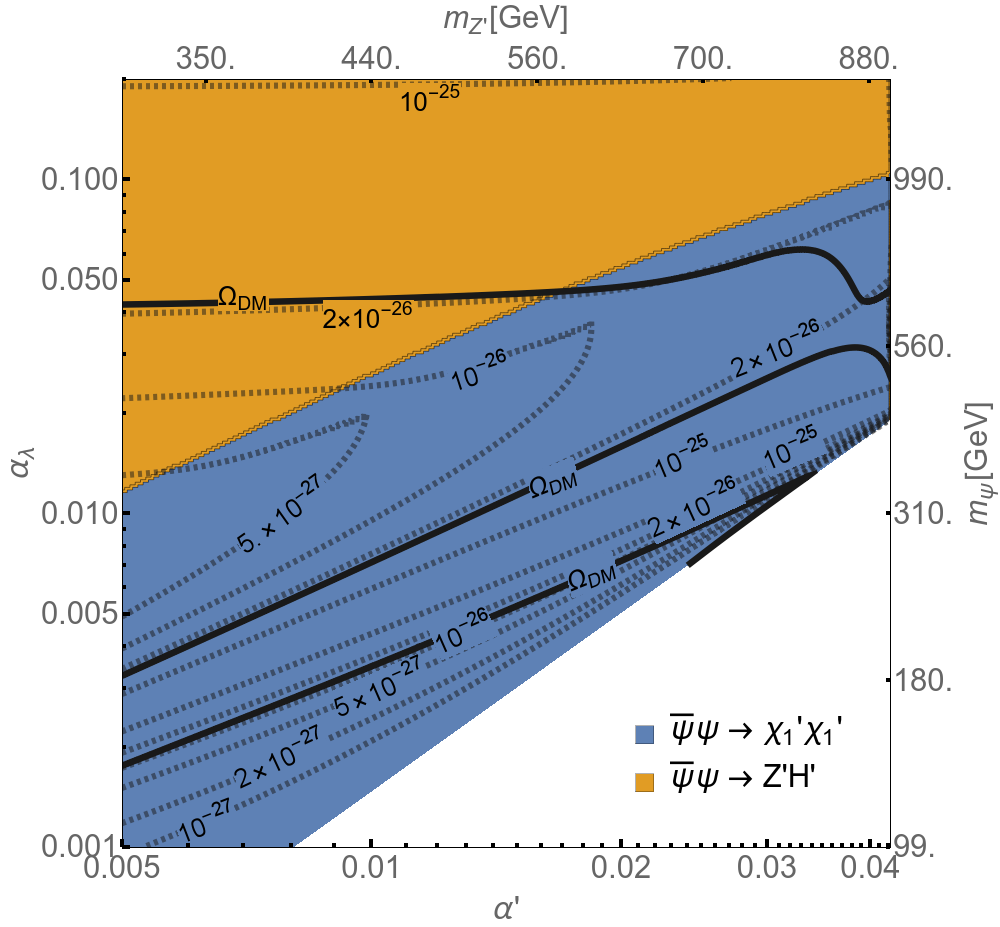}\\\textbf{b)$\langle \s v \rangle_{T=0} $}
	\end{minipage}
	
	\caption{\textbf{Left Panel (a):}
The black dashed contours give the dark matter relic density 
in units of the observed abundance $\Omega_{DM}h^2$;
the solid contour corresponds to points that produce the observed DM abundance. The dense set of contours in the $\chi_1' \chi_1'$ region corresponds to cases where the DM annihilates through the $Z^{\prime}$ resonance, $m_{\psi} \simeq m_{Z^{\prime}}/2$.  
\textbf{Right Panel (b):}  Contours denote T=0 annihilation cross sections $\langle \sigma v \rangle_0$, 
 relevant for indirect detection of dark matter.  In both panels, we have fixed $v'=1.25 $ TeV, $\tilde{m}_S=2.5$ TeV, $\tilde{m}_T=650$ GeV, $m_{\tilde{B}'}=1.5$ TeV, $\l A_{\lambda}=0.25$ TeV, and $\phi_{CP}=0$.}
	\label{fig:PsiAnnihilation}
\end{figure}

We now turn to a numerical treatment that encompasses more general cases with multiple channels and contributions. We use a combination of FeynRules \cite{Alloul:2013bka} and micrOMEGAs \cite{Belanger:2010pz,Belanger:2006is,Belanger:2001fz} with numerical diagonalization through ASperGe \cite{Alloul:2013fw}  to determine the relic abundance as well as the  $T=0$ cross sections relevant for indirect detection. Our results are shown in Figure \ref{fig:PsiAnnihilation} for a representative parameter set, and $\epsilon$ sufficiently small that all SM final states can be neglected. The color coding represents the strongest annihilation channel at each point of parameter space.  
The left panel shows that the $\chi'_1\chi'_1$ (blue) and $Z'H'$ (gold) channels tend to dominate on either side of the fixed point $\eta=1$.  
As $\alpha'$ increases, the lighter scalar mass decreases, owing to the presence of the D-term, see Eq.\,\ref{eqn:ScalarMassMatrix}. The right edge of the plot denotes $m_{S_1}=m_\psi$, beyond which the scalar is the LZP and the DM candidate. Close to this boundary, $\psi$ and $S_1$ are approximately degenerate, and coannihilation processes can dominate the freeze-out process (green region). The green region, corresponding to coannihilation into $\chi'_1 Z'$, features a resonant effect where the heavier neutralino $\chi_2'$ can go approximately on-shell; this occurs when  $m_{S_1} + m_{\psi} \approx 2 m_\psi \sim m_{\chi'_2}$.  In the upper-left corner of this plot, we expect relatively large loop corrections to the $H^{\prime}$ mass, which we have not included in our relic density calculations.  However, in this same region the hidden sector would also be fine-tuned for this set of supersymmetry breaking parameters, since the relatively light $Z^{\prime}$ would receive substantial corrections going like $\sim \lambda ^2 \tilde{m}_{S}^2$.  In addition, we do not expect the change in Higgs mass to have a significant impact on the relic density, since annihilation cross sections do not depend strongly on the Higgs mass in this region. 

In the right panel, we show dominant annihilation channels at $T=0$, along with (dashed) contours of the annihilation cross section. The solid contours denote regions with the correct relic density.  Again, over most of the parameter space, annihilations to $Z^{\prime} H^{\prime}$ dominate.  Along the upper solid contour, the dominant $p$-wave contribution (to $H^{\prime} H^{\prime}$) contributes a maximum of $\sim 3\%$ to the total annihilation cross section in the early Universe.  The result is that $\langle \sigma v \rangle_{T=0}$ is very nearly  the $s-$wave value of $2 \times 10^{-26}$ cm$^{3}$/s along this contour.

In either panel, annihilation rates to neutralino final states do not exceed those to $Z'H'$ when the latter channel is kinematically unsuppressed. Relative to the $Z^{\prime} H^{\prime}$ final state, whose tree-level cross section is $\propto \a_\l^{2}$,  $s$-channel contributions to neutralino final states are suppressed by powers of $\a' / \a_\l $ or $1/x_f$, with $x_f \equiv m_{DM}/T_{fo}$.  However,  neutralino final states are still relevant, if subdominant, where the scalar exchange diagrams are sufficiently large.   
This occurs near the right side of the plot, where a sizable $D$-term acts to suppress one of the scalar masses.  
Incidentally, 
the leading $s$-wave $\a_\l^{2}$ piece of neutralino diagrams is suppressed by the scalar mass splitting,  $m_{S_1}^2-m_{S_2}^2$. The $s$-wave annihilation rate to neutralinos assuming $\a' \ll \a_\l$ is
\begin{align}
	\langle \sigma v \rangle_{\chi_i' \chi_j'} &= C_i C_j (2-\delta_{ij})  \frac{\pi \a_\l^2  }{4 m_\psi^2}  \frac{m_\psi^4}{m_{S_1}^4} (2-s_{2\theta_S}^2)\frac{(1-m_{S_1}^2/ m_{S_2}^2)^2}{(1+m_{\psi}^2/m_{S_1}^2)^2(1+m_{\psi}^2/m_{S_2}^2)^2}\,,
\end{align}
where $C_1=\cos^2{\theta_N},\,C_2= \sin^2{\theta_N}$. This additional suppression can be understood by taking the $m_{S_1}^2 \rightarrow m_{S_2}^2$ limit and performing a Fierz transformation on the sum of the scalar mediated diagrams.  In this limit, these diagrams sum to give the operator $(\bar\psi \g_\m \g^5 \psi) \bar{\chi}_i \g^\m \g^5\chi_j$, which is helicity suppressed \cite{Kumar:2013iva}. In the limit $\a_\l \gg \a'$, the cross section for annihilation to neutralinos can reach exactly one-half the cross section of $Z'H'$, a limit saturated as $m_{S_2}/ m_{S_1} \to \infty$, $m_{S_1} \to m_\psi$, and $S_1 \to T^\ast  $ or $S$. For the explicit case shown in Fig. \ref{fig:PsiAnnihilation}, where $m_{S_2}/m_{S_1} \lesssim 6$, over a majority of the gold region annihilation to neutralinos contributes roughly $10-40 \% $ to the overall annihilation rate, with smooth interpolation to 1 at the fixed point line.

Figure \ref{fig:PsiAnnihilation} is only one slice of parameter space, and it is of interest to explore the dependence on other parameters ($\tilde{m}_{S,T}$, $v^{\prime}$, $m_{\tilde{B}'}$\,, and $A_{\lambda}$). The arguments of the previous paragraph summarize the dominant effect of varying the scalar soft masses -- they act as a dial that changes the relative importance of the neutralino final state(s) when both $Z^{\prime} H^{\prime}$ and neutralino final state(s) are kinematically accessible. 
The $m_{\tilde{B'}}$ chosen in the figure is such that the $\chi'_1\chi'_1$ final state can go on resonance, yet small enough that the heavier neutralino is still accessible.  A smaller $m_{\tilde{B}'}$ would alleviate some of the heavy neutralino kinematic suppression, allowing a marginally lighter thermal dark matter in regions where annihilations to the heavy neutralinos are relevant. Recall that for the dark matter to be a fermion, the trilinear $A_\l$ term must be small enough to not push a scalar mass below $m_\psi$. Otherwise, the primary effect of $A_{\lambda}$ is reflected via the impact of the scalar masses on annihilations to hidden neutralinos as described above.

Finally, the relic density is controlled by the overall mass scale.  In cases where an $s$-wave process dominates (typically the case here),  in the freeze-out approximation, $\Omega\sim m^2  \log(m)$, where $m$ is the mass scale associated with the annihilation cross section, $\sigma \propto m^{-2}$.   It is therefore possible to shift \textit{any} given contour to the correct DM abundance by rescaling (within the limits permitted by perturbativity considerations) all the mass scales in the hidden sector by the square root of the number displayed. When coannihilations become important, the scaling is still roughly $\Omega \propto m^{2}$, but with corrections that cause some deviation from this behavior (see \cite{Griest:1990kh} for further details). 

In summary, we have shown that fermionic dark matter with simple thermal histories is possible in our framework. In the majority of the parameter space, annihilations to $Z^{\prime} H^{\prime}$ provide for the ``hidden WIMP miracle," but more complicated pictures, including coannihilations or annihilations to hidden neutralinos, are possible. 

\section{Scalar Dark Matter}\label{sec:scalar}

In this section, we consider the scenario where the lighter scalar $S_1$ is the LZP dark matter candidate. For $S_1$ to be the LZP, one of two conditions are required: (i) one of the soft masses $\tilde{m}_S^2, \tilde{m}_T^2$ should be negative, or 
(ii) one of $\tilde{m}_S^2, \tilde{m}_T^2$ should not be too large (in this limit, SUSY relations force the scalar to be degenerate with the $\psi$), \emph{and} either the trilinear term $A_{\lambda}$ or $D$-term contribution 
 must be large enough to sufficiently split the eigenvalues to push the smaller eigenvalue below $m_{\psi}$. 
 
For case (i) with a single negative soft mass, in order to ensure open annihilation channels in the hidden sector, which is necessary to obtain the correct relic density via freeze-out (since we assume $\epsilon$ to be small), we must further have $2\a' <\a_\l$ or a seesawed down $m_{\chi_1'}$. In the limit of $\alpha_{\lambda} \gg \alpha'$ the cross sections are well approximated by 
 \begin{align}
\langle \sigma v \rangle_{S_1 S_1^\ast \to Z' Z'} &=\frac{\pi \a_\l^2 }{4 m_{S_1}^2} \left(  1 - \frac{|A_\l|^2}{m_{S_1}^2+m_{S_2}^2} \right)^2, \\
\langle \sigma v \rangle_{S_1 S_1^\ast \to H' H'} &=\frac{\pi \a_\l^2 }{4 m_{S_1}^2} \left(  1 - \frac{|A_\l|^2 \cos^2{2 \theta_S} }{m_{S_1}^2+m_{S_2}^2} -  
\frac{1}{2}\left( \frac{2m_\psi-|A_\l|\sin{2\theta_S} }{m_{S_1}} \right)^2 \right)^2 \label{eqn:ScalarToHH},\\
\langle \sigma v \rangle_{S_1 S_1^\ast \to \chi_i' \chi_j'} &= C_i C_j (2-\delta_{ij})\sin^2{2 \theta_S} \frac{2\pi \a_\l^2 }{ m_{S_1}^2}\left( \frac{m^2_{S_1}/m_\psi^2}{(1+m_{S_1}^2/m_\psi^2)^2} \right), \label{eqn:ScalarToNeutralinos}
\end{align}
where $C_i=\cos^2{\theta_N} ,\, C_2=\sin^2{\theta_N}$. Note that because the initial state is CP even, the relevant states are  $Z' Z'$  and $H^{\prime} H^{\prime}$, in contrast to the fermion case where $Z^{\prime} H^{\prime}$ played a starring role.  Of these states, $H'H'$ typically dominates when $m_\psi \gsim m_{S_1}$  on account of the  term that goes like   $m^2_\psi/m^2_{S_1}$ in Eq. (\ref{eqn:ScalarToHH}). This term is due to $t$- and $u$- channel diagrams generated via the $|T|^2|H'|^2$ term with one $H'$ set to its vev. 
Finally, the presence of $\sin{2 \theta_{S}}$ in  Eq. (\ref{eqn:ScalarToNeutralinos}) can be understood by noting that this channel receives a helicity suppression in the absence of scalar mixing. 

\begin{figure}[t]
	\centering
	\begin{minipage}{.5\textwidth}
	\includegraphics[width=1\columnwidth]{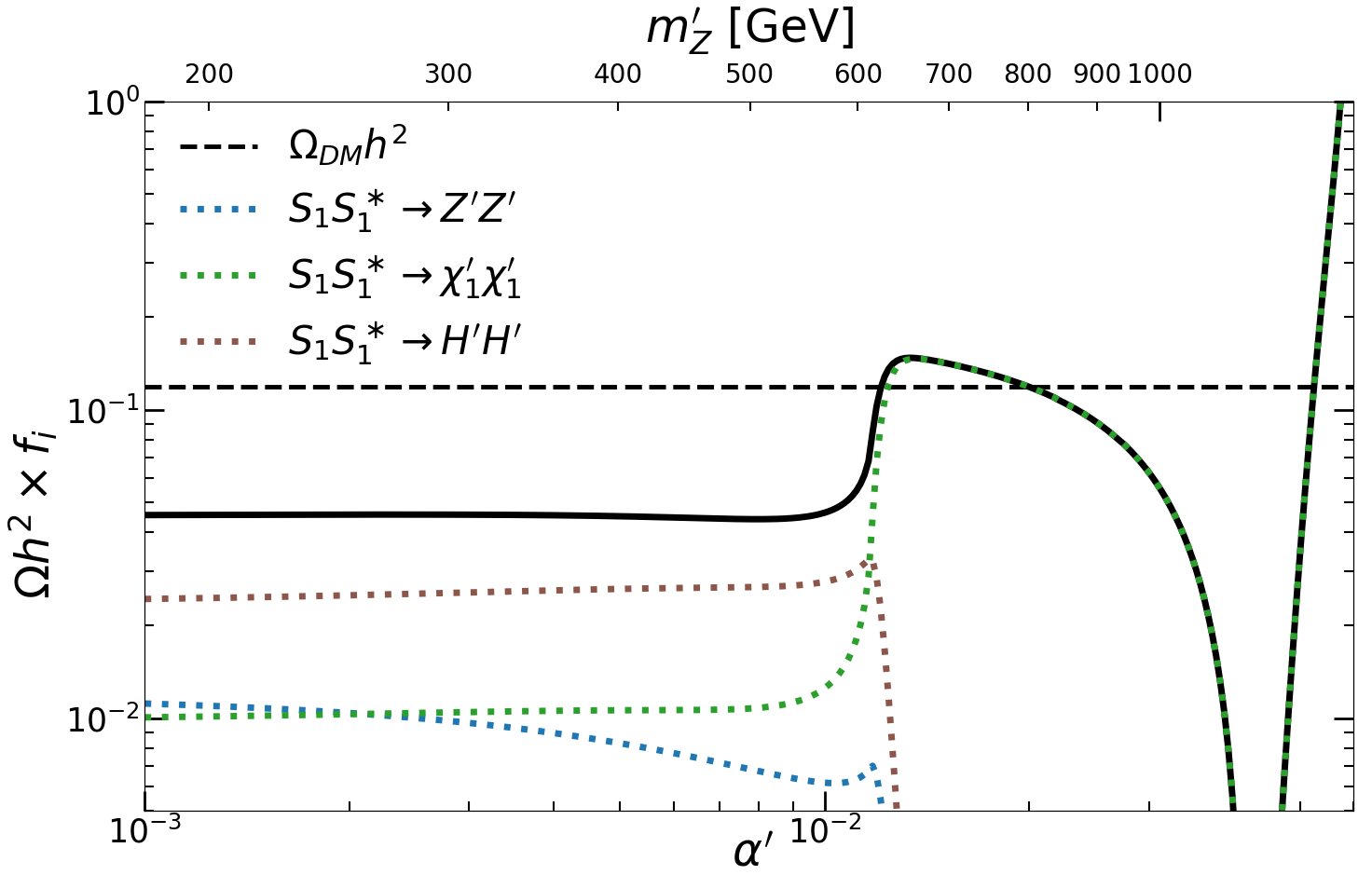}\\\textbf{a) $\Omega h^2 $}	\vspace{-0.11in}
	\end{minipage}%
	\begin{minipage}{.5\textwidth}\vspace{0.1in}
	\includegraphics[width=1\columnwidth]{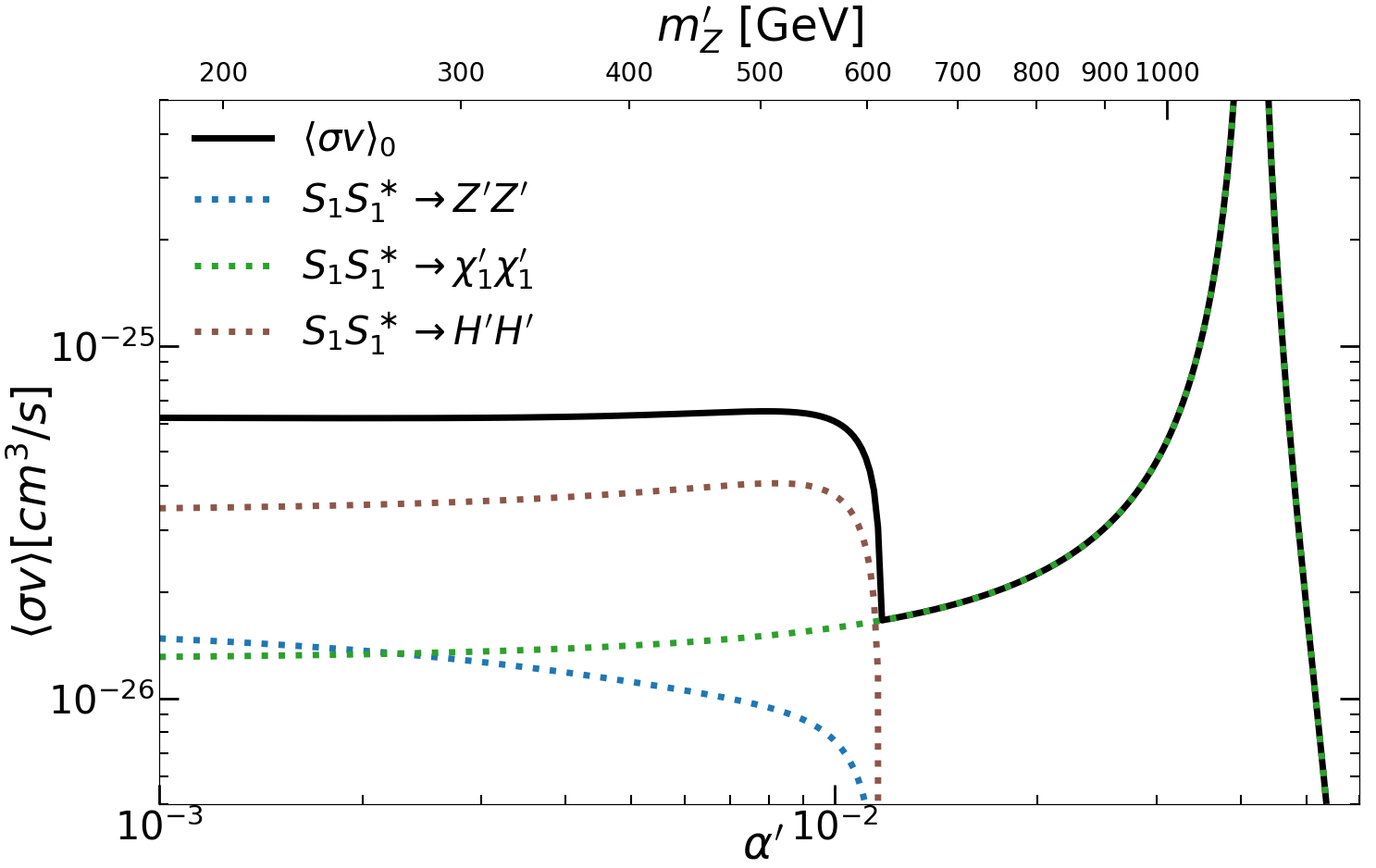}\\\textbf{b)$\langle \s v \rangle_{T=0} $}
	\end{minipage}
	
	\caption{ \textbf{Left Panel (a):} Relative contributions of various annihilation processes to $S_1$ freeze-out.  The relative importance is given by the Micromegas output $f_i\equiv\frac{\langle\sigma v\rangle_i}{\langle\sigma v\rangle_{\text{total}}}$ and corresponds to a freeze-out approximation for the annihilation rate of each channel \cite{Belanger:2001fz}.  \textbf{Right Panel (b):} The annihilation cross section  for $T=0$, relevant for indirect detection. For both panels, $\alpha_\l=0.045,\ v'=1.6$ TeV, $\tilde{m_S}^2=-400^2$ GeV$^2$, $\tilde{m_T}=1500$ GeV, $M_{\tilde{B}'}=3000$ GeV, $\l A_{\lambda}=600$ GeV, and $\phi_{CP}= -\pi$. }
	\label{fig:Scalar1DAnnhilation}
\end{figure}

Fig. \ref{fig:Scalar1DAnnhilation} shows the relative importance of these channels and illustrates a case where the correct relic abundance is realized via annihilation to neutralinos.   In the figure, $m_\psi=848$ GeV and  $m_{S_1}$ varies in the range $540-620$ GeV.  Even with this relatively modest hierarchy, scalars still dominantly annihilate to $H^{\prime} H^{\prime}$ when this channel is kinematically accessible.  In this case, and for relatively small $A$-terms, the dark matter abundance may be approximated as
\begin{align}
(\Omega_{S_1} h^2)_{H' H'} \approx \Omega_{DM} h^2 \left( \frac{0.12}{ \l }\right)^4 \left( \frac{2\, \text{TeV}}{m_\psi}\right)^4 \left( \frac{m_{S_1}}{500\, \text{GeV}}\right)^6\,.
\end{align} 
However, for large $\alpha^{\prime}$, this channel becomes kinematically inaccessible, and the relic density is set by annihilation into the only available channel in the hidden sector, $\chi_1^{\prime} \chi_1^{\prime}$. Although this channel is suppressed relative to the bosonic final states, Fig.  \ref{fig:Scalar1DAnnhilation} shows that it is still possible to achieve the correct DM abundance with this channel (even away from the $Z',\ H'$ pole).  

\begin{figure}[t]
	\centering
	\begin{minipage}{.5\textwidth}
	\includegraphics[width=1\columnwidth]{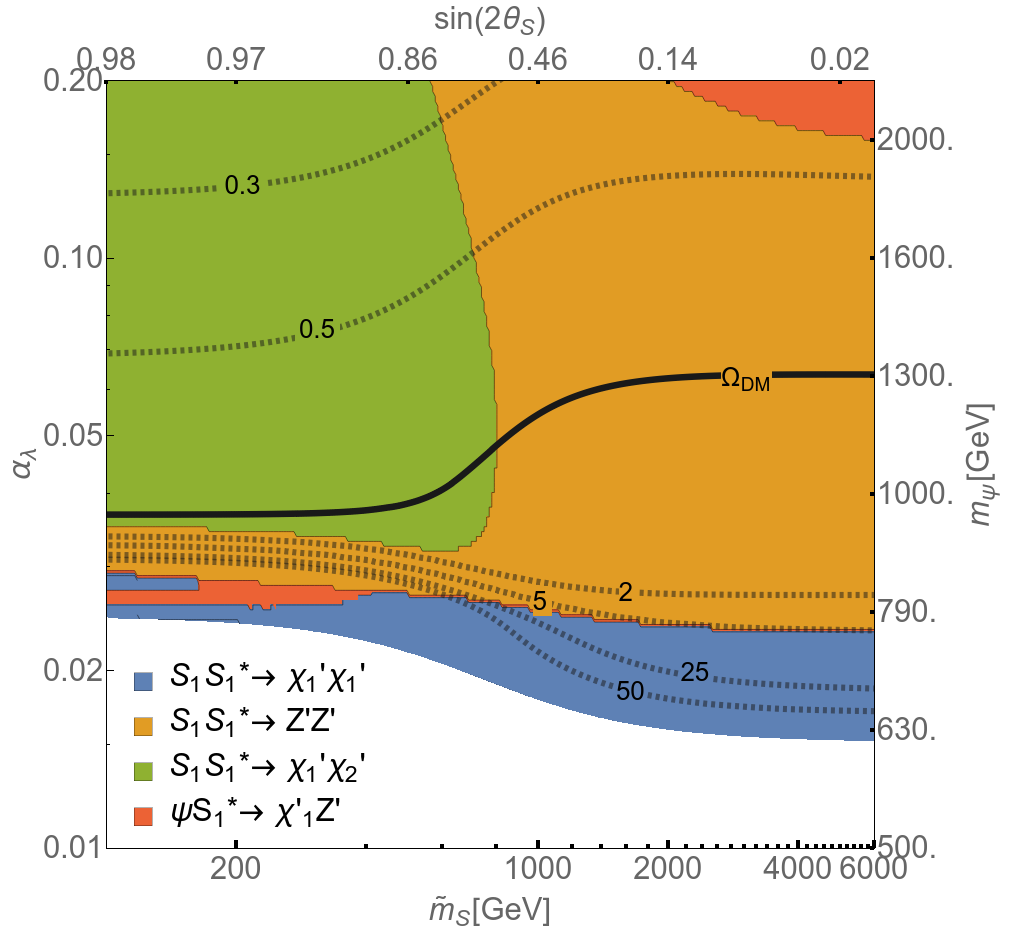}\\\textbf{a) $\Omega h^2/\Omega_{DM} h^2 $}
	\end{minipage}%
	~~
	\begin{minipage}{.5\textwidth}
	\includegraphics[width=1\columnwidth]{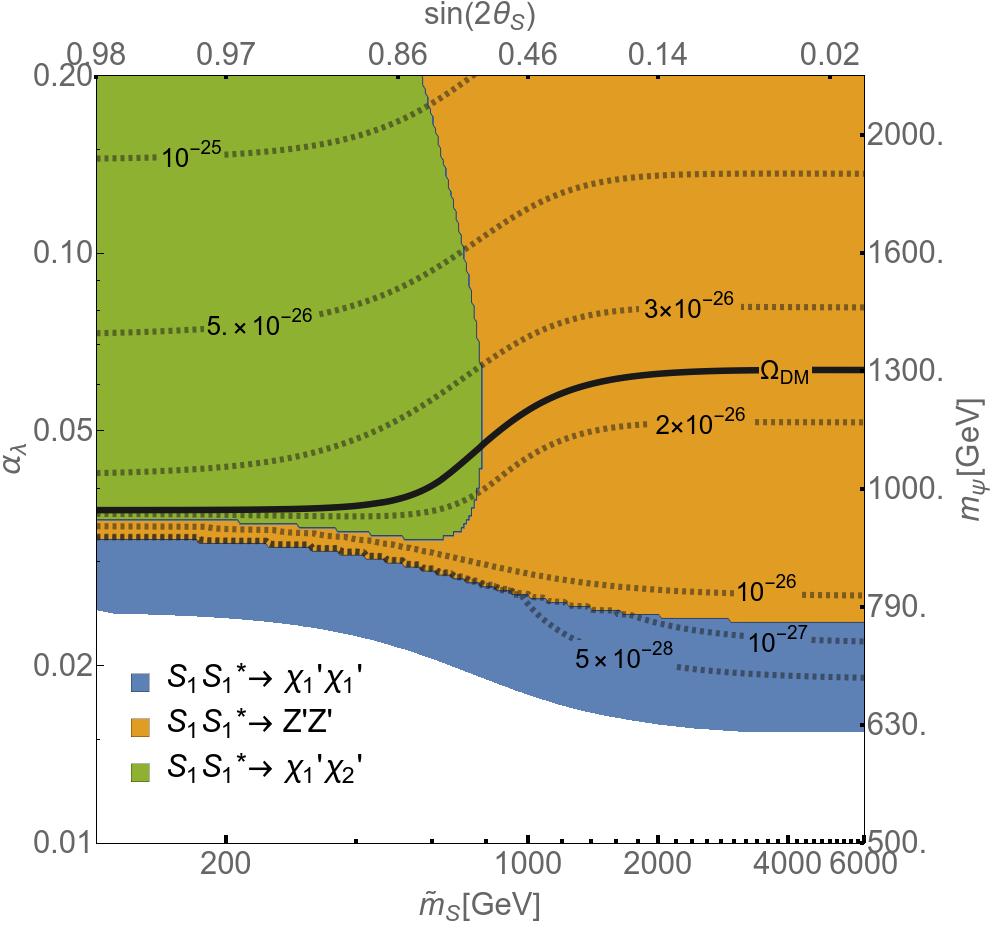}\\\textbf{b)$\langle \s v \rangle_{T=0} $}
	\end{minipage}
	
	\caption{ \textbf{Left Panel (a):} Color coding represents the dominant annihilation process contributing to $S_1$ freeze-out, while contours represent the computed DM relic abundance in units of the observed relic density.  \textbf{Right Panel (b):} The annihilation cross section  for $T=0$, relevant for indirect detection.  Again, different colors indicate the dominant annihilation channel. For both panels,  $\alpha'=0.01,\ v'=2$ TeV, $\tilde{m_T}=400$ GeV, $M_{\tilde{B}'}$=400 GeV, $\l A_{\lambda}=200$ GeV, and $\phi_{CP}$=0. }
	\label{fig:ScalarAnnihilation}
\end{figure}

 For case (ii), with positive soft masses, an interesting feature (for relatively modest $m_{\tilde{B}'}$) is a relatively compressed spectrum, with the gauge boson, Higgs boson, scalars, and fermions all in close proximity. 
The $A_\l$ term is not expected to be too large compared to the scalar masses, hence relatively compressed spectra are quite generic unless the $D$-term is very large. Such compressed spectra result in potentially richer cosmologies, including more robust possibilities of coannihilation. 

The interplay of the above processes can be seen in Fig. \ref{fig:ScalarAnnihilation}, where we plot the dominant annihilation processes over a slice of parameter space. At low $\a_\l$, the only kinematically accessible state for $S_1 S_1^\dagger$  annihilation is $\chi_1' \chi_1'$(blue region). However, in this blue region and for $\a_\l \gtrsim 0.02$, there is also an open coannihilation channel ($Z^{\prime} \chi_1^{\prime}$). This coannihilation is exponentially suppressed because the  $S_{1}$ --  $\psi$ mass splitting is still substantial, $\sim \mathcal{O}(100)$ GeV;
however, for a slice in the bottom left (red), a near perfect destructive interference amongst diagrams contributing to the $\chi_1'\chi_1'$ annihilation nevertheless allows for coannihilation to dominate. However,  because the overall annihilation rate is exceedingly small, the relic density far exceeds the observed relic density.  For larger $\a_\l$, annihilations to dark Higgs and gauge bosons dominate as they become accessible at $\a_\l\,>\,2\a'$ (gold region).  Eventually,  the $\chi_1' \chi_2'$ (green region) state becomes kinematically accessible and marginally exceeds these channels. The neutralino channels diminish as we move to the right side of either panel, owing to the decrease in $\sin {2 \theta_{S}}$; see the discussion surrounding  Eq.~(\ref{eqn:ScalarToNeutralinos}).  For sufficiently small $\theta_S$, the process is driven by otherwise subdominant pieces suppressed by  $g'/\l$, which are not shown in Eq.~(\ref{eqn:ScalarToNeutralinos}).   Note that, again,  the largest values of $\alpha_{\lambda}$ shown here correspond to tuned hidden sectors, particularly for the largest values of $m_{S}$.

The fact that $Z'Z'$ exceeds $H'H'$ here is a consequence of the relatively degenerate spectrum; the enhancement that hidden Higgs final states receive from factors of $m_\psi/m_{S_1}$, discussed below Eq.~(\ref{eqn:ScalarToHH}),
is no longer substantial. For example, in the bottom right corner when $Z'Z'$ goes on-shell, $m_\psi/m_{S_1}\sim 1.09$.   This permits terms proportional to $g^{\prime}$ (neglected in  Eq.~(\ref{eqn:ScalarToHH})) to allow annihilation to $Z^{\prime} Z^{\prime}$ to dominate. The $S_1$ and $\psi$ grow closer in mass towards the top right, and coannihilation processes are seen to become important (red region). As discussed in the previous section, any of the contours can be made to match the correct relic density by rescaling the mass scales involved. 

We see from these figures that unlike when $\psi$ is the LZP, the case where $S_1$ is the LZP is more involved, with several available annihilation processes that are viable candidates for setting the relic abundance. The link between indirect detection and freeze out, shown by comparing the left and right panels of Figs. \ref{fig:Scalar1DAnnhilation} and \ref{fig:ScalarAnnihilation}, is, however, relatively straightforward due to the nearly universal presence of $s$-wave processes. Exceptions occur in regions where coannhilation processes dominate, rendering a suppressed indirect detection signal.

\section{Coannihilation Regime}\label{sec:coannihilations}
The regime where the mass gap between the LZP and its superpartner is small is worthy of special attention. As seen in earlier sections, in this regime coannihilations between the two can be important for setting the dark matter relic density. Moreover, the small mass gap can cause the heavier of the two to have a long lifetime, which can have important cosmological and phenomenological consequences.  

The heavier state decays to its superpartner and a trio of SM fermions via an off-shell neutralino and the RPV coupling through a dimension-7 operator,  which we write schematically as
\begin{equation}
{\mathcal O}_{decay} = \frac{( S_1  \, \psi) (  \psi_{SM} \psi_{SM} \psi_{SM})}{\Lambda^3},
\end{equation}
where $\Lambda$ is a combination of gaugino and sfermion masses, and $\psi_{SM}$ represents a SM fermion. The identity of the fermions depends on the texture of RPV couplings but does not affect our discussion here.  The important effect is that the large power of $\Lambda$ in the denominator, coupled with the phase space suppression due to the small mass splitting and the 4-body final state, can lead to an extremely long decay lifetime. Assuming $\psi$ to be lighter without loss of generality, the decay width for $S_1\to \psi\psi_{SM} \psi_{SM} \psi_{SM}$ is schematically
\begin{equation}
\Gamma\sim\frac{\epsilon^2g'^2g_Y^2\lambda''^2}{2^9\pi^5}\frac{\Delta m^7}{m_0^4 m_{LSP}^2},
\end{equation}
where $\Delta m=m_{S_1}-m_{\psi}$, and $m_{0}$ represents a generic scalar superpartner mass in the visible sector. This decay width corresponds to a lifetime
\begin{equation}
\tau/s\sim\left(\frac{10^{-4}}{\epsilon g' \lambda''}\right)^2\left(\frac{10\,\text{GeV}}{\Delta m}\right)^7\left(\frac{m_0}{\text{TeV}}\right)^4 \left(\frac{m_{LSP}}{\text{TeV}}\right)^2.
\end{equation}
A lifetime $\tau\gsim 1$\,s could disrupt Big Bang Nucleosynthesis (BBN) through late injection of energetic photons and charged fermions.    This potentially imposes a strong constraint on $\Delta m$ and hence coannihilation as a viable method to produce the correct dark matter relic density, though the strength of the constraint depends on other unknown parameters. One possibility that allows for sufficiently short lifetimes -- even in the presence of small $\Delta m$ -- is for $\epsilon$ to be fairly large, which would  have interesting implications for direct detection.

Another possibility, following  \cite{Poulin:2016anj,Dienes:2018yoq}, is to note that  particle decays after BBN are allowed so long as the energy injected into photons and $e^\pm$ is several orders of magnitude smaller than the energy density in dark matter. This can indeed be the case. Understanding the constraints requires an understanding of the energy density stored in $S_{1}$. For instance, suppose the coannihilation process that sets the $\psi$ dark matter relic abundance is $\psi S_1\to \chi_1'Z'$.  In this case, the cross-processes $S_1 Z' \to \psi\chi_1'$ and $S_1\chi_1' \to \psi Z'$ can continue to deplete the $S_1$ abundance until they freeze out at a later time. Assuming $m_\psi\approx m_{S_1}$, we can approximate the energy density in $S_1$ relative to the energy density in $\psi$ DM after $S_1$ freeze-out as 
\begin{equation}
\frac{\rho_{S_1}}{\rho_\psi}\sim \text{Exp}\left[-\frac{m_\psi}{T_{fo}}\left(\frac{m_{\psi}}{\text{min}(m_{\chi_1'},Z')}-1\right)\right],
\end{equation}
where $T_{fo}$ is the DM freeze-out temperature. Due to this exponential dependence, we expect lifetimes with post-BBN decays to be compatible for $\frac{m_{\psi}}{\text{min}(m_{\chi_1'},Z')} \gsim 1.4$ even if all of the $S_1$ energy density were ultimately converted into photons and $e^\pm$. Note, however, that much of the $S_1$ energy goes into LZP dark matter, and only a small fraction $\sim \Delta m/ m_{S_1}$ goes into photons and $e^\pm$, significantly mitigating such constraints. Furthermore, given that $n_{S_1} \ll n_{DM}$ at the time of dark matter freeze-out, we also do not expect subsequent $S_1$ scattering or decay processes to contribute a significant additional population of dark matter. 

\section{UV considerations}\label{sec:UV}

In the previous sections, we illustrated several incarnations of the WIMP miracle in the hidden sector that differed both in the identity of the dark matter and the most important annihilation channel.  In this section, we examine whether and how these various scenarios arise from reasonable choices of parameters at a high scale, and whether these are compatible with constraints on weak scale MSSM parameters.  

We pay particular attention to what mass scales are reasonable in the hidden sector under the assumption that supersymmetry breaking is communicated similarly to the two sectors, as might occur with gravity mediation. 
One must renormalization group (RG) evolve the resulting parameters from the scale at which SUSY breaking is mediated to the weak scale, relevant for dark matter phenomenology.  
Evolution of the hidden sector parameters is performed with $\beta$-functions from Ref.\,\cite{Andreas:2011in} with slight corrections (see Appendix for details). 
A linear combination of soft masses convenient for RG evolution, which also serves as a rough proxy for the overall scale of the hidden sector, is
\begin{equation}
\Sigma \equiv \frac{1}{3}(\tilde{m}_T^{2}+\tilde{m}_S^{2}+\tilde{m}_{H^{\prime}}^{2}).
\label{eqn:Sigma}
\end{equation}
The physical mass spectrum is related through $(m_{S_1}^2+m_{S_2}^2)/2 = $Tr($m_{scalar}^2)/2=\frac{2}{3}\Sigma+m_\psi^2$. 
When evolved into the IR, which for concreteness we evaluate at the top quark mass, we find the following approximate expression, derived from numerical RGE flow graphs shown in the Appendix, Fig. \ref{fig:IR_Sigma-scale}:\begin{equation}
 \Sigma(m_{top}) = (0.3,\ 1) \Sigma_0 + (0,\ 0.4) m_{\tilde{B}^{\prime}_0}^2.
 \end{equation}
Here $m_{\tilde{B}^{\prime}_0}$ is the value of the soft mass of the hidden gaugino at the high scale. The lower (upper) boundary of the range of the $\Sigma_0$ coefficient  corresponds to large (small) $\a_\l$, whereas the lower (upper) boundary for the range of the  $m_{\tilde{B}'_0}^2$ coefficient  corresponds to small (large) $\a'$. Just as the large top Yukawa in the MSSM suppresses stop masses in the IR,  large $\lambda$ can suppress  the dark scalar masses in the IR. And as in the (no-scale) MSSM, the gaugino mass can generate scalar masses at one-loop.  However, because the abelian dark $U(1)^{\prime}$ runs to weak coupling in the IR, this suppresses the IR gaugino mass and mitigates its effects on the scalar masses.

To understand what we expect for the hidden sector mass scale in Eq.~(\ref{eqn:Sigma}), we should compare to the parameters of the MSSM, whose values we have some indirect clues about from the LHC. We make use of the approximate solutions to MSSM RGEs from, e.g.,  \cite{Carena:1996km,Kazakov:2000us,Drees:2004jm,Hikasa}.  
For $\tan{ \beta} =10$, IR SUSY breaking parameters are related to universal boundary conditions for scalars ($m_0$), gauginos ($m_{1/2}$), and trilinears ($A_{0}$) as:
\begin{eqnarray}
\label{eqn:MSSMUVIR}
\tilde{m}_{Q_3}^2 &\approx &0.63 m_0^2+ 5.7 m_{1/2}^2 - 0.13 m_{1/2} A_{0},\\
\tilde{m}^2_{U_3}  &\approx&  0.26 m_0^2+ 4.3 m_{1/2}^2  - 0.27 m_{1/2} A_0 ,\\
\tilde{m}_{H_2}^2  &\approx & -0.12 m_0^2 - 2.6 m_{1/2}^2  -0.40m_{1/2} A_0, \\
(M_3,\ M_2,\ M_1)&\approx&(2.9,\ 0.82,\ 0.41)m_{1/2}.
\end{eqnarray}
We are agnostic about the precise UV SUSY-breaking boundary conditions, but universal boundary conditions such as these enable us to get a sense of the rough scales involved.

Some of these parameters are constrained by LHC data. In particular, direct searches constrain stops and gluinos to be TeV scale or heavier.   But it is also possible to say more.  Given the absence of a definitive hint of the mass scale of superpartners, we can take the measured mass of the Higgs boson, $m_h=125$ GeV, as an indirect measure of the stop mass scale. It is  known that this Higgs mass is compatible with stop masses below a TeV in the presence of significant stop mixing, but LHC direct searches place this  scenario in tension. If mixing in the stop sector is not near maximal, then the observed Higgs mass requires that stops be much heavier, $\gsim 5$ TeV \cite{Giudice:2011cg,Draper:2013oza,Bagnaschi:2014rsa,Vega:2015fna}. From the equations above, we see that the stop mass in the IR is largely determined by two UV mass parameters: a soft mass scale $m_0$ and the gaugino mass scale $m_{1/2}$.  The gaugino mass piece provides the dominant contribution in the IR unless $m_0\gg m_{1/2}$. In the gaugino mass dominated scenario, $\sim5$ TeV stops suggest $m_{1/2}, m_0\sim$ TeV. Assuming that $\Sigma_0, m_{\tilde{B}'_0}$ are comparable to their respective MSSM counterparts $m_0^2$ and $m_{1/2}$ in the UV, hidden sector RG running then suggests $\mathcal{O}(100)$ GeV - $\mathcal{O}$(TeV) as the mass scale for hidden sector particles. Scalar mass dominated scenarios are correlated with somewhat heavier hidden sector masses. Therefore, $\mathcal{O}(100)$ GeV - $\mathcal{O}$(TeV) scale hidden sector particles can be generically compatible with multi-TeV stops and gluinos in the MSSM sector without any significant mass hierarchies in the UV.  Detailed information on the hidden sector spectrum requires RG evolution of the splittings between the various soft masses. The splittings decrease fairly slowly and so are sensitive to their UV starting points.

\begin{figure}[t]
\includegraphics[width=0.9\columnwidth]{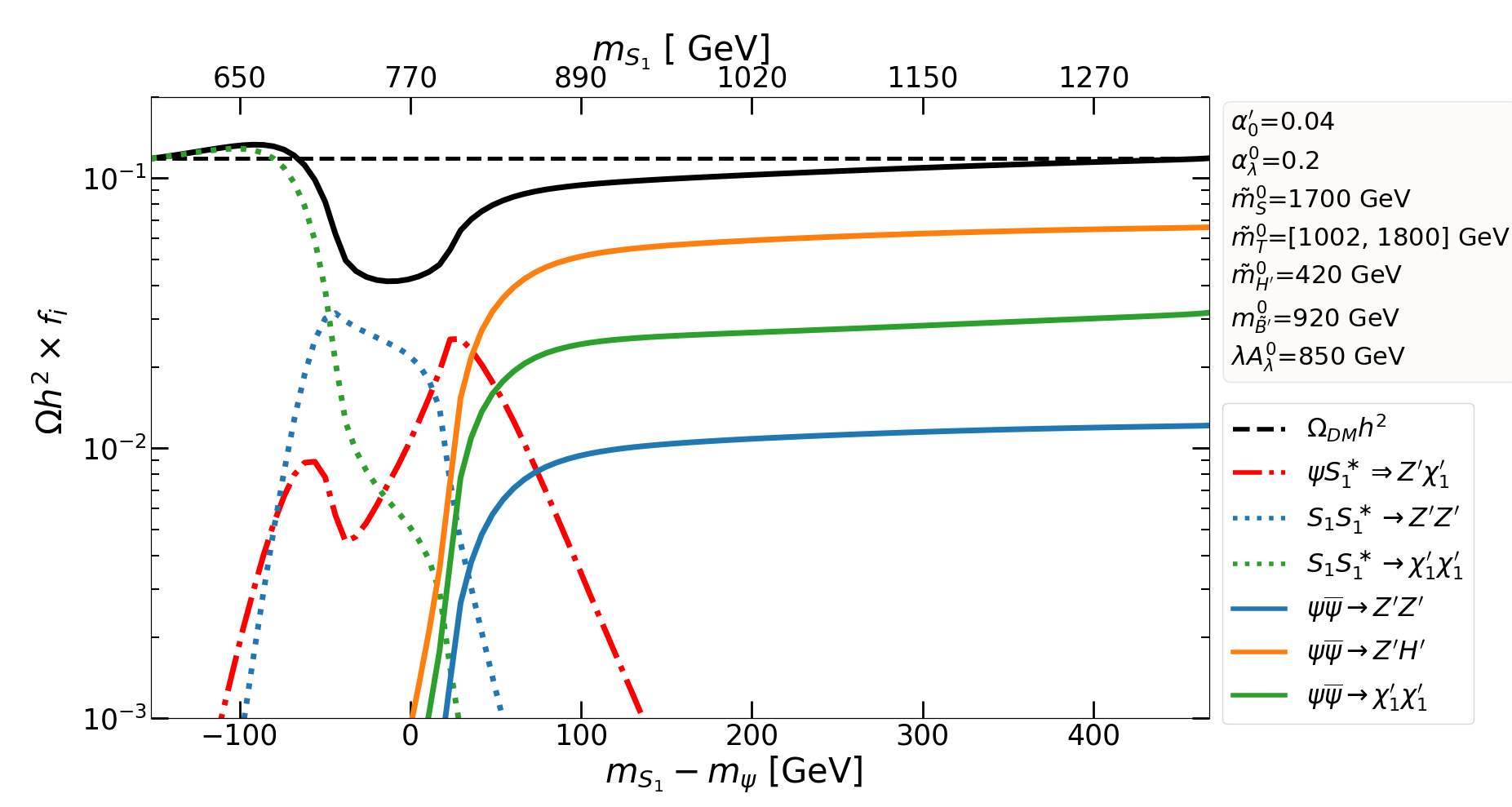}
\caption{Interpolation between dark matter scenarios by scanning the soft mass $\tilde{m}_{T}^0$ in the UV.  The solid black curve indicates the realized dark matter density, while the other curves indicate the relative importance of different annihilation channels, with $f_i$ as defined in the caption of Fig.\,\ref{fig:Scalar1DAnnhilation}. The parameters in the legend are defined in the UV. }
\label{fig: UVinterpolate}
\end{figure}

We now turn from these generalities to make a more firm connection with the cosmological histories outlined in earlier sections. We use analytic one-loop formulae for IR quantities (see Appendix) to plot relic abundances given UV initial conditions. In Fig.~\ref{fig: UVinterpolate}, we show (as a black curve) the realized relic abundance as a function of the IR mass splitting between the lightest scalar $S_{1}$ and the fermion $\psi$ in the hidden sector.  This splitting depends on $\tilde{m}_{T}^0$ in the UV.  The various curves denote the relative importance of various annihilation channels -- solid (dotted) curves for fermion (scalar) dark matter -- as well as coannihilation (dashed curves). 
At large $m_{S_1}-m_\psi$, we can see that the correct relic abundance is realized for a wide range of mass splittings, consistent with thermal histories dominated by $\psi \bar{\psi} \rightarrow Z^{\prime} H^{\prime}$ annihilations (solid orange curve). Annihilations to the lighter neutralino are smaller by $\mathcal{O}(1)$ factors and therefore not negligible (solid green curve).  On the left hand edge of the plot, where scalar dark matter is realized, $S_{1} S_{1}^{\ast} \rightarrow \chi_1'\chi_1'$ annihilation dominates (dotted green). When the mass splitting is small, coannihilations can dominate (dot-dashed curve). This figure therefore shows that for a range of TeV scale input parameters in the UV, the correct relic abundance can be realized for both fermion and scalar dark matter scenarios. It is worth pointing out that at the right edge of the plot, $S_1$ transitions from being dominantly $\tilde{T}^{\ast}$ to $\tilde{S}$.  For sufficiently large $\tilde{m}_T$,  $\tilde{m}_S$ will be pushed down via the impact of $\tilde{m}_T$ on the RG flow of $\tilde{m}_S$, so further increase of $\tilde{m}_T$ actually results in a decrease in the lightest scalar mass.
 
Finally, we comment briefly on the evolution of the dimensionless couplings. There is a $\lambda=\sqrt{2} g'$ IR fixed point, which  can be understood as the emergence of an $\mathcal{N}=2$ SUSY, where $\psi$ is degenerate with $Z',\,H'$. The distance from this fixed point, $r$, has a simple solution at one loop
\begin{equation}
\label{eqn:rEvolution}
r(t) \equiv 1-\frac{2\alpha'(t)}{\alpha_\lambda(t)}\,,   \quad r(t) = r_0 \left(\frac{\alpha'(t)}{\alpha_0'}\right)^3=
\frac{r_{0}}{\left(1-\frac{\alpha_0' t}{\pi}\right)^3}\,,
\end{equation}
where $t\equiv \text{log}(\mu/M_{GUT})$,\, $M_{GUT}=2 \times 10^{16} \text{ GeV}$, and null subscripts correspond to GUT boundary conditions. Interestingly, $r(t)$ is also a measure of the kinematic suppression of $\psi 
\bar{\psi}$ annihilation to $Z', H'$ (see Fig. \ref{fig:kinematic_Suppression} and surrounding text in the Appendix). This equation thus indicates the possibility of kinematic suppression of this channel as a consequence of RG evolution. 

\section{Decay modes of Hidden Sector Particles}\label{sec:decays}

In this section we discuss the decay modes of various hidden sector particles. This is crucial for indirect detection, as once the dark matter annihilates into these particles, their decay modes will determine the spectra of SM states that will be observed by experiments.

\subsection{$H'$ decays}

Kinetic mixing induces a mixed quartic interaction between the visible and hidden sector Higgs fields via a D-term contribution to the potential shown in Eq.~(\ref{eq:Dterm}). This generates a mixed mass matrix after each field acquires a vev. In the $(H_d^0,\ H_u^0,\ H^{\prime})$ basis, this is (to leading order in $\epsilon$)
\begin{equation}
m_{H_d,H_u,H'}^2 = 
\begin{pmatrix}
s^2_{\beta}\, m_A^2 + c^2_{\beta}\, m_Z^2 & -s_{\beta} c_{\beta}(m_A^2+m_Z^2) & \epsilon s_{\theta_W} m_Z m_{H^{\prime}} c_{\beta} \\
-s_{\beta} c_{\beta}(m_A^2+m_Z^2) & c^2_{\beta}\, m_A^2 + s^2_{\beta}\, m_Z^2 + \delta/s^2_\beta & -\epsilon s_{\theta_W} m_Z m_{H^{\prime}} s_{\beta} \\
\epsilon s_{\theta_W} m_Z m_{H^{\prime}} c_{\beta} & -\epsilon s_{\theta_W} m_Z m_{H^{\prime}} s_{\beta} & m_{H'}^2
\end{pmatrix},
\end{equation}
where we have used the abbreviations $s_x = \sin{x}$ and $c_x = \cos{x}$, and tan$\beta=\frac{v_u}{v_d}$ is the ratio of the up- and down-type Higgs vevs, with $v_u^2+v_d^2=v^2=(246 ~\text{GeV})^2$. The $\delta$ term encodes the radiative contribution to the Higgs mass, which we adjust to recover the 125 GeV Higgs mass. After rotating the visible sector Higgs fields by the standard MSSM Higgs mixing angle $\alpha$, the mass matrix in the $(H,\ h,\ H^{\prime})$ basis is
\begin{equation}
m_{H,h,H'}^2 = 
\begin{pmatrix}
m_{H}^2 & 0 & \epsilon s_{\theta_W} m_Z m_{H^{\prime}} c_{\alpha+\beta} \\
0 & m_{h}^2 & -\epsilon s_{\theta_W} m_Z m_{H^{\prime}} s_{\alpha+\beta} \\
\epsilon s_{\theta_W} m_Z m_{H^{\prime}} c_{\alpha+\beta} & -\epsilon s_{\theta_W} m_Z m_{H^{\prime}}  s_{\alpha+\beta} & m_{H'}^2
\end{pmatrix}
.
\end{equation}
In the MSSM decoupling limit, $m_A \gg m_Z$, $\alpha \approx \beta-\frac{\pi}{2}$.  If $H$ is so massive that it decouples from this system,\footnote{One must take care in taking this strict decoupling limit. For processes such as the decay $H'\to \chi_1 \chi_1$,  the contribution via $H_u$ may be suppressed relative to those from $H_d$, for instance due to $\chi_1$ having a roughly $\tan{\beta}$ larger content of $H_d \sim H$ than $H_u \sim h$.  In this case, effective decoupling can be delayed.} we get 
\begin{equation}
m_{h,H'}^2 = \left(
\begin{array}{cc}
m_{h}^2 & \epsilon s_{\theta_W} m_Z m_{H^{\prime}} c_{2\beta} \\
\epsilon s_{\theta_W} m_Z m_{H^{\prime}} c_{2\beta} & m_{H^{\prime}}^2
\end{array}
\right).
\end{equation}
The mass eigenstates of this matrix are comprised of the $h, H'$ states with mixing angle 
$\theta_H$ approximately given by
\begin{equation}\theta_H \approx -\frac{\epsilon s_{\theta_W} m_Z m_{H^{\prime}} c_{2\beta}}{m_{H^{\prime}}^2-m_{h}^2}.\label{eqn:thetaH}
\end{equation}

To understand which decays are allowed for $H'$ requires an understanding of the $H'$ mass relative to those of other hidden sector particles.  As discussed in Sec.~\ref{sec:framework}, we expect an approximate degeneracy between $H'$ and $Z^{\prime}$ to be maintained even after accounting for loop corrections.
This eliminates the possibility of the decay channel $H^{\prime} \rightarrow Z^{\prime} Z^{\prime}$, except in extremely fine-tuned regions (for moderate fine-tuning, it might be possible that $Z Z^{\prime}$ could be open).  More likely is the possibility of $H^{\prime}$ decays to neutralinos.   While the hidden sector neutralinos are also degenerate with the $H^{\prime}$ in the supersymmetric limit, recall that a somewhat large $m_{\tilde{B'}}$ produces a seesaw effect that makes (the mostly $\tilde{H'}$) $\chi_1'$ light, opening the channel $H^{\prime} \rightarrow \chi'_{1} \chi'_{1}$. The width for this channel is  
\begin{equation}
\Gamma(H^{\prime} \rightarrow \chi_1^{\prime}  \chi_1^{\prime} ) = \frac{g'^2 m_{H^\prime}}{4 \pi}(\sin^2\theta_N \cos^2\theta_N) \left(1-\frac{4m_{\chi_1^{\prime}}^2}{m_{H^{\prime}}^2}\right)^{3/2},
\end{equation}
where $\theta_N$ is the $\tilde{B}^{\prime}-\tilde{H}^{\prime}$ mixing angle, see Eq.~(\ref{eqn:neumat}).

If this decay channel is not kinematically accessible, the $H'$ decays into SM states with an $\epsilon^2$ suppression. Because $H'$ inherits the couplings of the SM Higgs via $\theta_H$ mixing, it may decay into SM states such as $WW,\, ZZ,\, t\bar{t}$ or $hh$ \cite{Schabinger:2005ei}, 
or to visible sector superpartners, especially neutralinos. Decays to $hh$ are also directly mediated via the Higgs portal coupling. Decays to MSSM Higgs states, e.g., $AA, HH, H^+H^-$, are possible but likely kinematically suppressed, and we do not consider them further for simplicity. The final possibility is the decay into neutralinos of both sectors, $H' \rightarrow \chi_{1}^{\prime} \chi_{1}$, which occurs at the same order in $\epsilon$.

Among the visible sector SM states, $H'\to WW$ will dominate so long as $m_{H'} > 2 m_W$.  In the approximation given by Eq.~(\ref{eqn:thetaH}), the partial widths to SM bosons are:
\begin{eqnarray}
\Gamma(H^{\prime} \rightarrow W W ) &=& \frac{\epsilon^2 g_Y^2 c^2_{2\beta} m_{H'}}{64 \pi}\left(\frac{1-4(m_W/m_{H'})^2+12(m_W/m_{H'})^4}{1- m_{h}^2/ m_{H'}^2}\right) \sqrt{1-\frac{4 m_{W}^2}{m_{H'}^2}}, \nonumber \\
\Gamma(H^{\prime} \rightarrow Z  Z ) &=& \frac{\epsilon^2 g_Y^2 c^2_{2\beta} m_{H'}}{128 \pi}\left(\frac{1-4(m_Z/m_{H'})^2+12(m_Z/m_{H'})^4}{1- m_{h}^2/ m_{H'}^2}\right) \sqrt{1-\frac{4 m_{Z}^2}{m_{H'}^2}},  \\
\Gamma(H^{\prime} \rightarrow h  h ) &=& \frac{\epsilon^2 g_Y^2 c^2_{2\beta} m_{H'}}{128 \pi}\left(1+\frac{3 c^2_{2\beta} m_Z^2}{ m_{H'}^2- m_{h}^2}\right)^2 \sqrt{1-\frac{4 m_{h}^2}{m_{H'}^2}}. \nonumber
\end{eqnarray}
In the large $m_{H'}$ limit, we get $\Gamma(H^{\prime} \rightarrow W W ) \approx 2 \Gamma(H^{\prime} \rightarrow Z  Z ) \approx 2 \Gamma(H^{\prime} \rightarrow h h)$
as expected from the Goldstone equivalence theorem. 

The decay width into MSSM neutralinos $H'\to\chi_i\chi_j$ is subdominant to the above widths due to the relatively small Yukawa coupling suppressed by $\frac{m_Z}{\mu}.$
The decay into the neutralino combination $H^{\prime} \rightarrow \chi_1 \chi_1^{\prime}$, on the other hand, can dominate in some regions of parameter space.
This process is generated by neutralino mixing. After diagonalizing the kinetic terms of the neutralinos via $\tilde{B} \to \tilde{B} - \epsilon \tilde{B}' $, the mass matrix in the basis $\tilde{\chi}^{\prime}_{\text{gauge}}=\left( \tilde{H}' \  \tilde{B}'  | \tilde{B} \ \tilde{H}_D \  \tilde{H}_U  \right)$ is (to leading order in $\epsilon$)
\setlength{\tabcolsep}{6pt} 
\renewcommand{\arraystretch}{1} 
\begin{equation}
m_{\chi}=
\left(
\begin{array}{cc|ccc}
   0 & g' v' & 0 & 0 & 0  \\
 g' v' & m_{\tilde{B}'}& \epsilon  m_{\tilde{B} \tilde{B}'}-\epsilon  m_{\tilde{B}} &  \frac{1}{2} g_Y \epsilon  v_d & -\frac{1}{2} g_Y \epsilon  v_u \\
 \hline
 0 & \epsilon  m_{\tilde{B} \tilde{B}'}-\epsilon  m_{\tilde{B}} & m_{\tilde{B}} & -\frac{1}{2} g_Y v_d & \frac{1}{2}g_Y v_u \\
 0 & \frac{1}{2} g_Y \epsilon  v_d &  -\frac{1}{2} g_Y v_d &  0 & -\mu   \\
 0 & -\frac{1}{2} g_Y \epsilon  v_u & \frac{1}{2}g_Y v_u & -\mu  & 0  \\
\end{array}
\right).\label{eqn:gaugino6x6}
\end{equation}
Here, we assume that the wino is sufficiently heavy to be decoupled from the analysis. Diagonalizing the neutralino and Higgs mass matrices as $m_{\chi,\,\text{diag}}= N m_{\chi} N^\dagger$\,, $m_{\text{diag}}^2= U m^2_{H_d,H_u,H'} U^\dagger$, we calculate the relevant decay width as
\begin{align}
	\Gamma(H'\to\chi _1 \chi _1')=&\frac{g^{\prime 2} m_{H'}}{8 \pi } \sqrt{\left(1-\frac{m_{\chi_1}^2}{m_{H'}^2}-\frac{m_{\chi_1^\prime}^2}{m_{H'}^2}\right)^2-\frac{4 m_{\chi_1}^2 m_{\chi_1^\prime}^2}{m_{H'}^4}} 
	\left(1 -\frac{m_{\chi_1}^2+m_{\chi_1^\prime}^2+2 m_{\chi_1} m_{\chi_1^\prime}}{m_{H'}^2}\right) 
 \notag \\ &  \times
   \left(  U_{1',H'}(N_{1,\tilde{B}'} N_{1',\tilde{H}'} +  N_{1,\tilde{H}'} N_{1',\tilde{B}'})  \right)^2 .
\end{align}

\begin{figure}
	\begin{minipage}{.45\textwidth}
	\includegraphics[width=1\columnwidth]{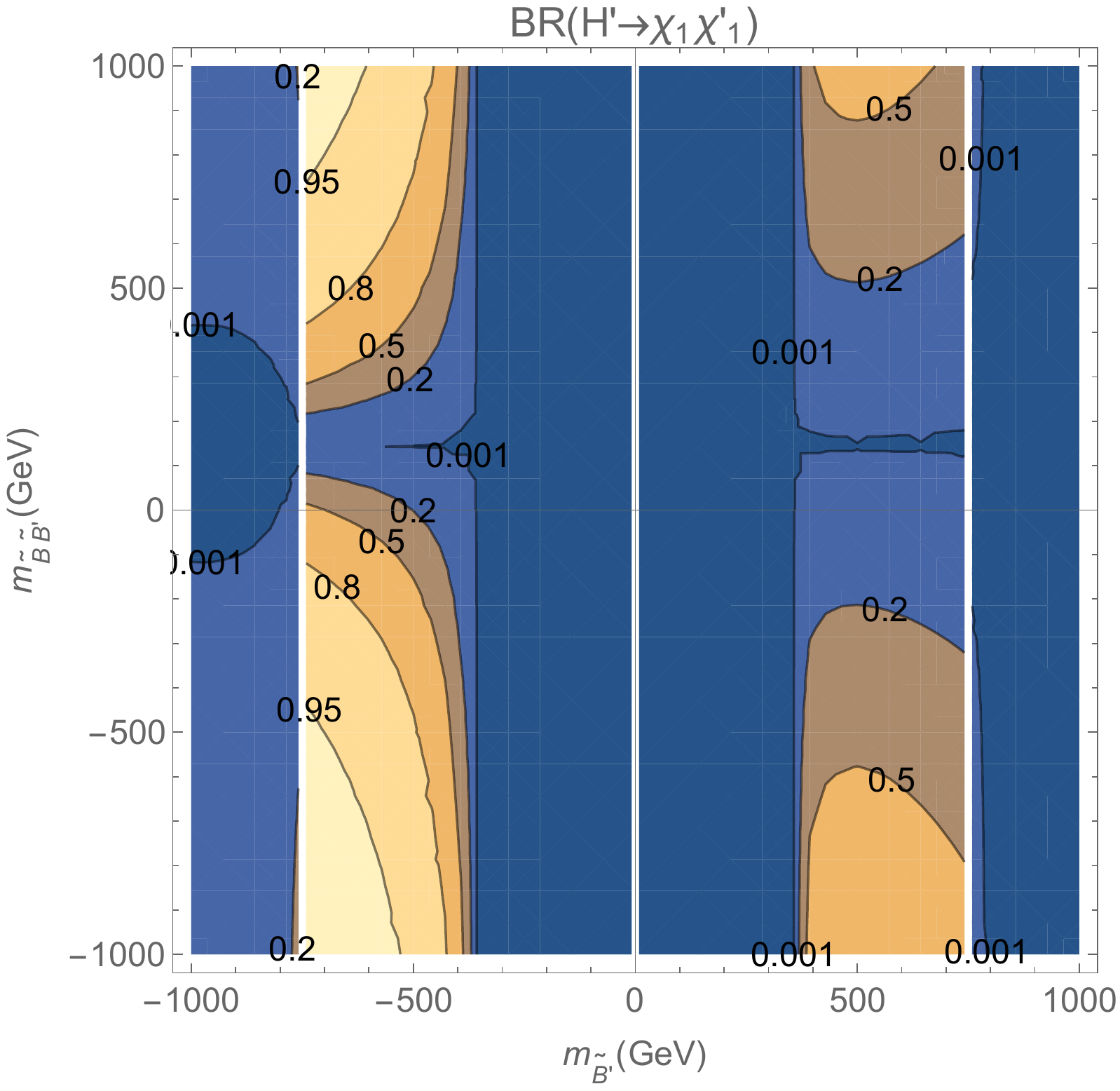}
	\end{minipage}%
	~~
	\begin{minipage}{.55\textwidth}\vspace{0in}
	\includegraphics[width=1\columnwidth]{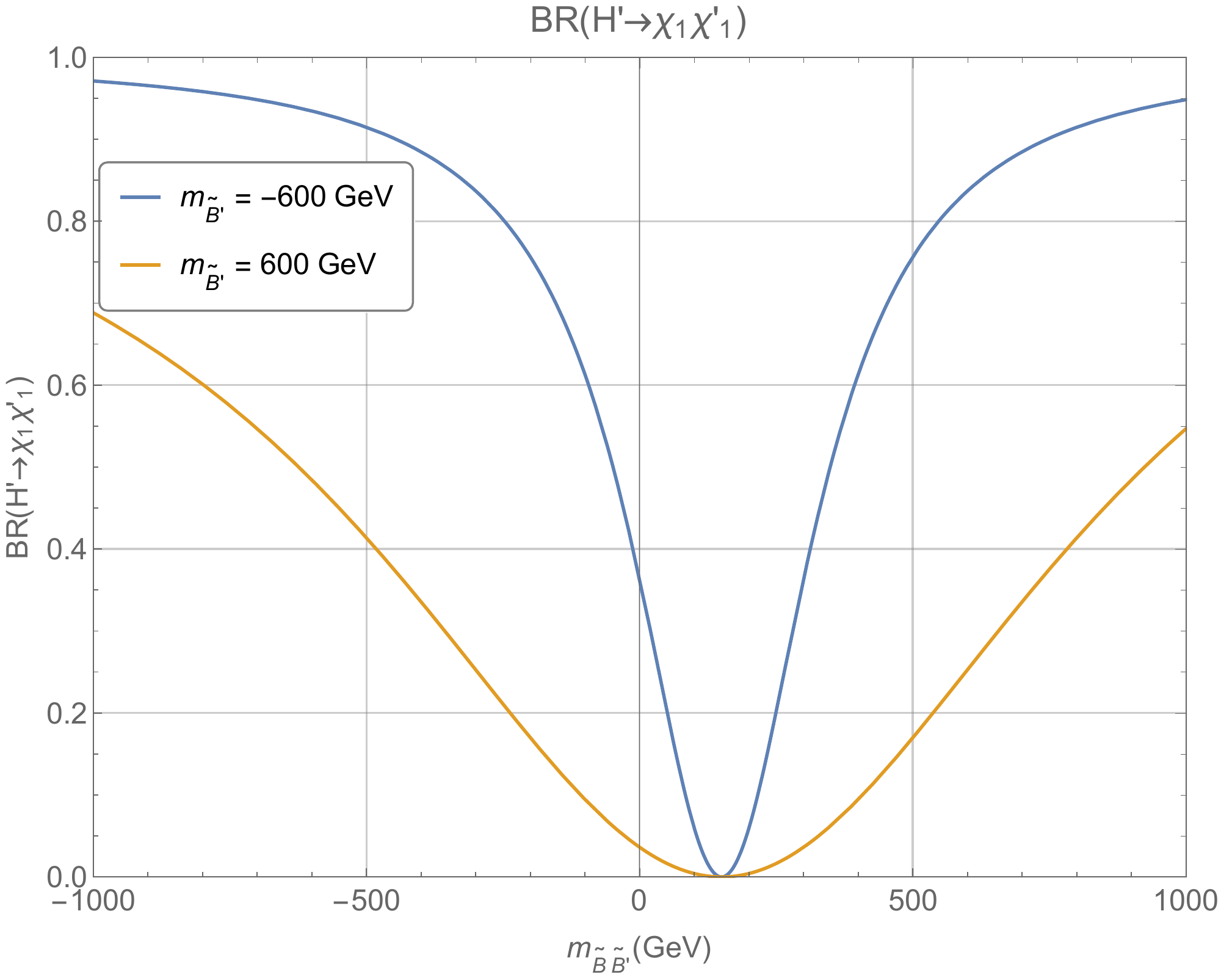}
	\end{minipage}
\caption{Branching ratio of the hidden Higgs boson to the lightest MSSM and hidden neutralino, BR($H^{\prime} \rightarrow \chi_1 \chi_1^{\prime})$, as contours in the hidden gaugino mass $m_{\tilde{B'}}$ - mixed gaugino mass $m_{\tilde{B} \tilde{B}^{\prime}}$ plane (left panel) and for $m_{\tilde{B'}}= 600, - 600$ GeV (right panel). The other parameters are set to $\tan\,\beta=10$, $m_{\tilde{B}}=150$ GeV, $\mu=1000$ GeV, $\epsilon=.01$, $g'=1$, and $m_{H'}=500$ GeV.}
\label{fig:HiggsContour}
\end{figure}
The neutralino masses are allowed to be negative in this formula, and the first index of $N_{ij},\, U_{ij}$ denotes mass eigenstates, with $i=1^{\prime}, 2^{\prime},1,2, 3$, with the prime indicating that the eigenstate is dominantly comprised of hidden sector fields.  
The second index indicates states after diagonalizing kinetic terms but prior to mass diagonalization. 
In Fig.~\ref{fig:HiggsContour}, we explore the branching ratio into this decay channel in the $H$ decoupling limit and to leading order in $\epsilon$.  In the contour plot in the left panel, the branching ratio vanishes for small $|m_{\tilde{B'}}|$ because the channel becomes kinematically inaccessible,  $m_{\chi}+m_{\chi'}>m_{H'}$.  At  large $|m_{\tilde{B'}}|$ it is negligible because the hidden sector decay $H^{\prime} \rightarrow \chi'_{1} \chi'_{1}$ becomes kinematically accessible and dominates.  In between, large mixing between the hidden neutralino and the lightest MSSM neutralino can make this mixed channel dominant, reaching branching ratios over 90\%. The right panel shows two slices of this contour plot at $m_{\tilde{B'}}= 600, - 600$ GeV. This plot illustrates that the relevant branching ratio can vanish for some value of $m_{\tilde{B} \tilde{B}^{\prime}}$ where contributions from field redefinition to remove the kinetic mixing of Eq.~(\ref{eq:Kmixing}) and the diagonalization to remove the mass mixing introduced in Eq.~(\ref{eqn:gauginomassmix}) conspire to cancel each other. For $\chi_1 \sim \tilde{B}$ (equivalently, $m_Z \ll \mu $) and small $\epsilon$, this cancellation occurs when $m_{\tilde{B}} \approx m_{\tilde{B} \tilde{B}'}$, as seen from the relevant off-diagonal term in Eq.~(\ref{eqn:gaugino6x6}).  

In summary, we expect $H^{\prime}$ decays to be dominated by $H^{\prime} \rightarrow WW$, $H^{\prime} \rightarrow \chi_{1}^{\prime} \chi_{1}^{\prime}$, $H^{\prime} \rightarrow \chi_{1}^{\prime} \chi_{1}$, or $H^{\prime} \rightarrow Z^{\prime}Z$. Decays to hidden sector particles will be followed by cascades into SM final states.

\subsection{$Z'$ decays}
The hidden sector decay $Z^{\prime} \rightarrow \chi_1'\chi_1'$ will dominate if kinematically accessible since all other channels are $\epsilon$ suppressed; otherwise, decays to pairs of SM fermions or to $\chi_1 \chi_1^{\prime}$ dominate. In the $m_{Z'}\gg m_Z$ limit, the couplings of the $Z^{\prime}$ to fermions are simply proportional to hypercharge as induced by the kinetic mixing, and the residual change to the coupling coming from the diagonalization of the $Z-Z^{\prime}$ mass matrix is negligible. The $Z^{\prime}$ dominantly decays to up-type quarks, followed closely by charged leptons.  In the opposite $m_{Z'}\ll m_Z$ limit, the $Z'$ instead primarily couples to electric charge, again decaying dominantly to up-type quarks (except the top quark, which is now kinematically inaccessible) or charged leptons.  Decays to $WW$ or $hZ$ are small, at the $10^{-4}$ level or below, for $m_{Z^\prime}  < 100$ TeV.  

\begin{figure}
\includegraphics[width=0.6\textwidth]{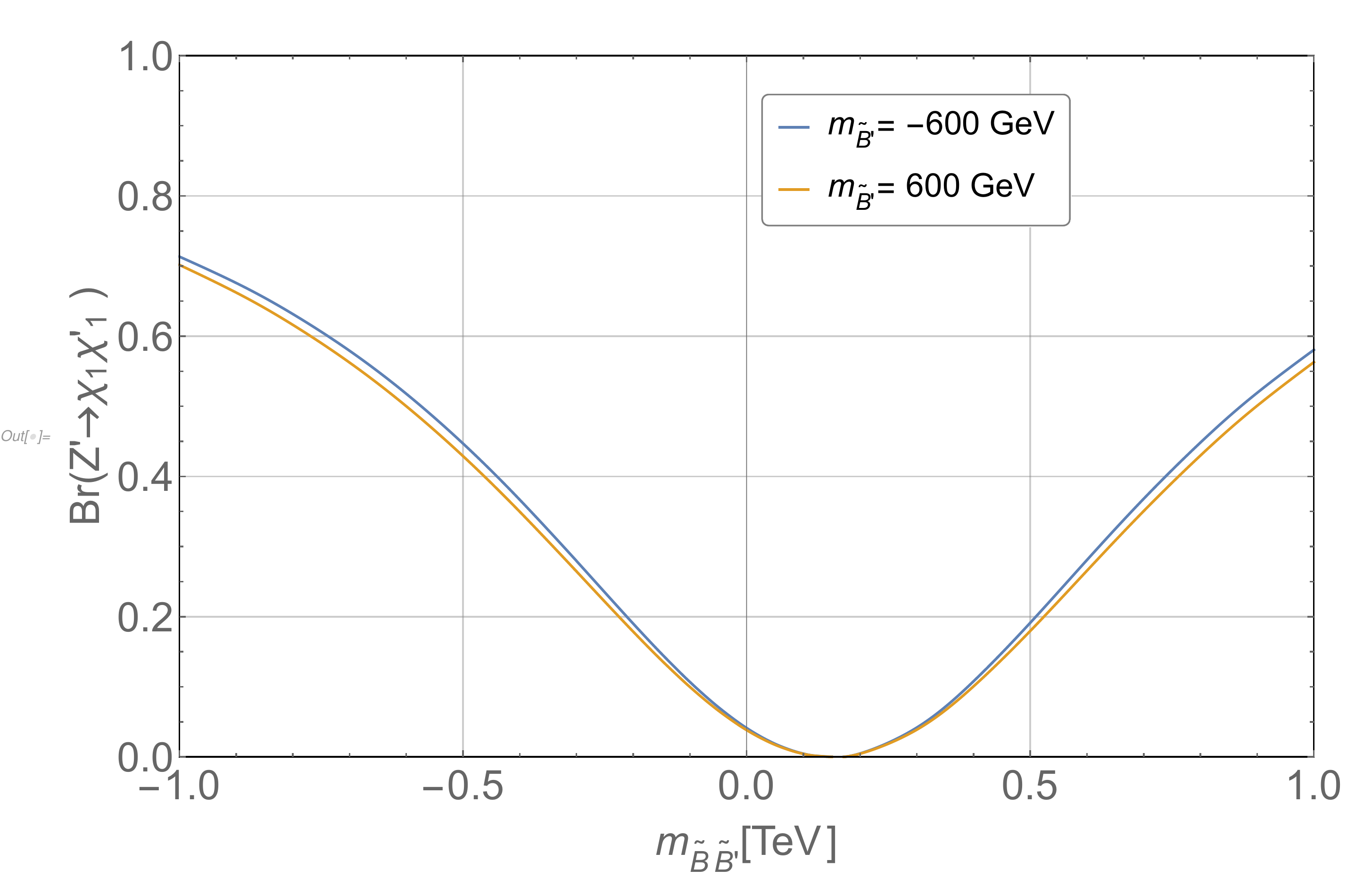}
\caption{Branching ratio of the hidden gauge boson to the lightest MSSM and hidden neutralino, BR($Z^{\prime} \rightarrow \chi_1 \chi_1^{\prime})$, for $m_{\tilde{B'}}= 600, - 600$ GeV. The other parameters are set as in Fig.~\ref{fig:HiggsContour}: $\tan\,\beta=10$, $m_{\tilde{B}}=150$ GeV, $\mu=1000$ GeV, $\epsilon=.01$, $g'=1$, and $m_{H'}=500$ GeV.}
\label{fig:ZContour}
\end{figure}

For $Z'\to\chi_1 \chi_1^{\prime}$ decay to dominate requires a larger neutralino mass mixing between the two sectors than in the Higgs boson case, see Fig.~\ref{fig:ZContour}.  This is because unlike the  $H'$, which couples to the gaugino-Higgsino combination in neutralinos, the $Z'$ couples to the Higgsino-Higgsino combination. Since the $\chi_1^{\prime}$ is mostly $\tilde{H^{\prime}}$ and the $\epsilon$-suppressed coupling to $\chi_1$ is via the $\tilde{B'}$ component, the $Z^{\prime} \rightarrow \chi_1 \chi_1^{\prime}$ coupling suffers from an additional $\tilde{H'}-\tilde{B'}$ mixing angle suppression relative to the $H^{\prime} \rightarrow \chi_1 \chi_1^{\prime}$ coupling.  
The relative closeness of the two curves for different signs of $m_{\tilde{B}}$ compared to the $H^{\prime}$ decay case (Fig.~\ref{fig:HiggsContour} right panel) is merely an artifact of our choice of parameters, and can be modified by changing $\theta_{N}$.

\subsection{$\chi_1'$ decays}
We are interested in scenarios where $\chi_1'$ is the lightest fermion in the hidden sector; hence all of its decays are into the visible sector via the portal coupling and are $\epsilon$ suppressed. The decay must proceed through the gaugino component of $\chi_1'$, denoted by $N_{1'1}$. In the limit where R-parity is unbroken, $\chi_1'$ decays into the LSP $\chi_1$ via an off-shell sfermion, with the width 
\cite{Pierce:2019ozl, Arcadi:2015ffa}
\begin{equation}
\Gamma (\chi_1'\to \chi_1 f\bar{f})=\frac{\epsilon^2\alpha_Y^2 N_{1'1}^2}{64\pi}\frac{m_{\chi_1'}^5}{m_0^4}\,f_2(m_{\chi_1}^2/{m^2_{\chi_1'}}),
\end{equation}
where $f_2(x)=1-8x+8x^3-x^4-12x^2 \log{x}$, and $m_0$ is the sfermion mass scale. If kinematically allowed, it can also decay as $\chi_1'\to\chi_1 (h/Z)$ through the bino-Higgsino mixing in the visible sector (if the $\chi_1'-\chi_1$ splitting is smaller than the $h$/$Z$ mass, the boson can be off-shell, giving a 3-body decay). This on-shell decay channel is subdominant to the above channel if $|\mu| > \frac{8\pi}{g_Y} \left(\frac{m_0}{m_{\chi_1'}}\right)^2 m_Z$. If R-parity violation is significant, $\chi_1'$ inherits the RPV decay channel of $\chi_1$ into three SM fermions: \begin{equation}
\Gamma (\chi_1'\to udd+\bar{u}\bar{d}\bar{d})=\frac{3\,\epsilon^2\lambda''^2 N_{1'1}^2\alpha_Y}{128\pi^2}\frac{m_{\chi_1'}^5}{m_0^4}.
\end{equation}
For sufficiently large $\lambda''$, this can be the dominant decay channel for $\chi_1'$. 
 
\section{Direct Detection and Collider Constraints}\label{sec:direct}

Direct detection in scenarios of hidden sector dark matter through the kinetic mixing portal has previously been studied by, e.g., \cite{Evans:2017kti,Chun:2010ve}. The spin-independent cross section per nucleon can be written to leading order in $\epsilon$ as:
\begin{equation}
\sigma = \frac{C \epsilon^2 g'^2 e^2 Z^2 c^2_{\theta_W} \mu_{D n}^2}{\pi A^2 m_{Z'}^4},
\end{equation}
where $\mu_{D n}$ is the reduced mass of the dark matter particle and the nucleon, $Z$ is the atomic number, $A$ is the mass number, and $C=\{ \frac{1}{4},  \, \cos^4\theta_S \}$ for \{fermion, scalar\} dark matter,  with $\theta_S$ the mixing angle between $S_1$ and $S_2$. Because $m_Z$ is much greater than the momentum exchange in the scattering process, the cross section goes like coupling to the electromagnetic current $e^2 Z^2$.

\begin{figure}
\includegraphics[width=12cm]{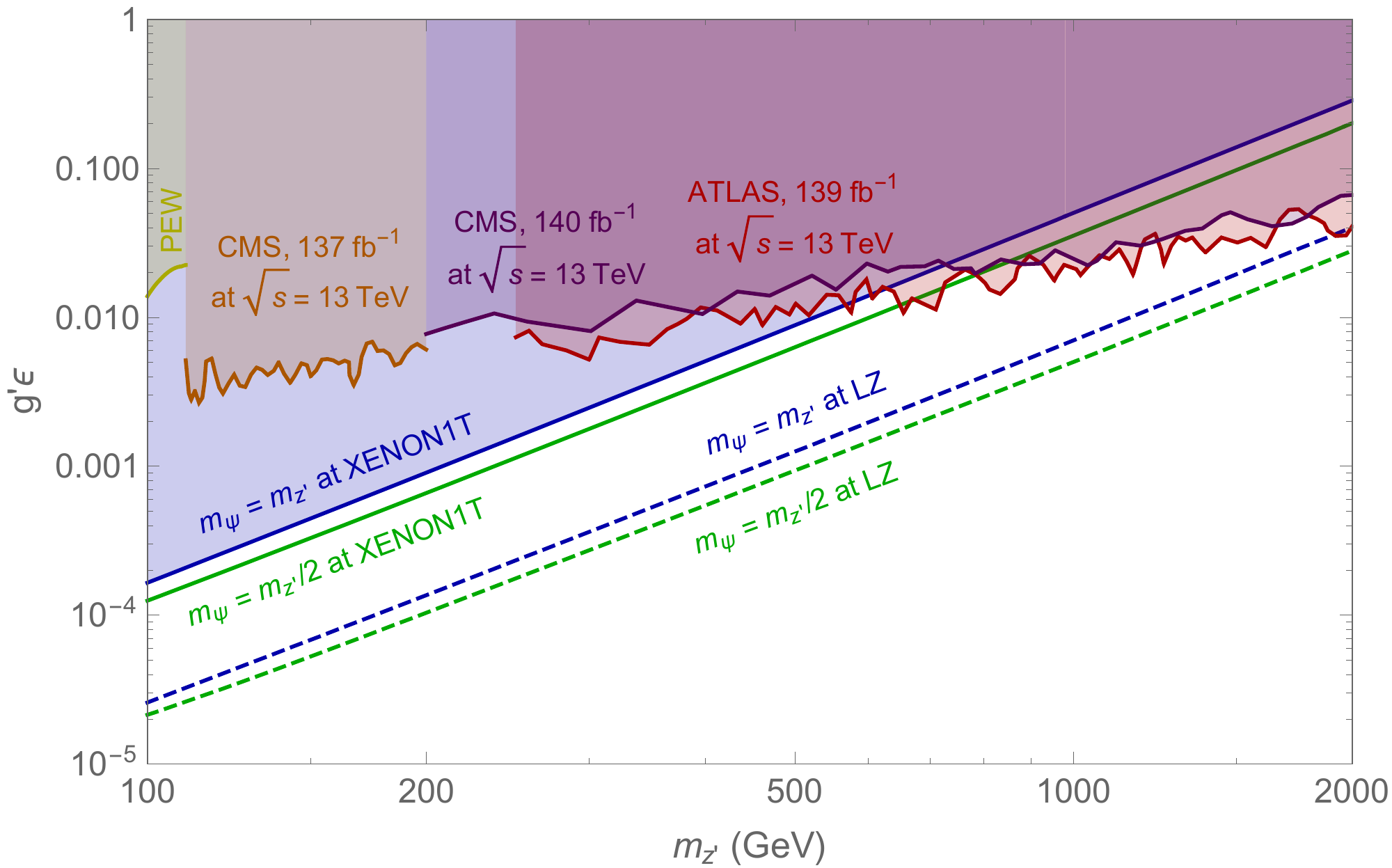}
\caption{Constraints on the product $g'\epsilon$ from direct detection for fermion dark matter for $m_{\psi} = m_{Z'}$ (blue) and $m_{\psi} = \frac{m_{Z'}}{2}$ (green).  Solid lines denote constraints from XENON1T, and dotted lines denote projected constraints from LZ after 1000 live days with a 5.6 tonne fiducial mass.   LHC constraints are plotted assuming $g'=1$, as derived from CMS searches for a resonance decaying to muon pairs \cite{Sirunyan:2019wqq} for $110$ GeV $< m_{Z^{\prime}} < 200$ GeV (orange), from CMS searches for dilepton resonances \cite{CMS:2019tbu} for $m_{Z^{\prime}}>200$ GeV (purple), and from ATLAS searches for dilepton resonances \cite{Aad:2019fac} for $m_{Z^{\prime}}>250$ GeV (red).  Precision electroweak constraints \cite{Evans:2017kti} are plotted assuming $g'=1$ (yellow).} 
\label{fig:Direct Detection}
\end{figure}

Given these cross sections, we can derive constraints on the product $g'\epsilon$ from direct detection experiments.  Current constraints from XENON1T \cite{Aprile:2018dbl} and projected constraints from LZ \cite{Akerib:2018dfk} are shown in  Fig.~\ref{fig:Direct Detection}.  We plot two curves for each experiment, assuming $m_{\psi}=m_{Z'}$ or $m_{\psi} = \frac{m_{Z'}}{2}$ for fermion dark matter. The choices are representative of different parameter regimes that replicate the correct relic abundance: The choice $m_{\psi} = m_{Z'}$ is inspired by the IR fixed point $\lambda=\sqrt{2} g'$, whereas $m_{\psi} = \frac{m_{Z'}}{2}$ represents the region of parameter space where the annihilation $\psi \bar{\psi} \rightarrow \chi^{\prime}_i \chi^{\prime}_j$ occurs through a $Z'$ resonance (note that small variations around the $m_{\psi} = \frac{m_{Z'}}{2}$ resonance can precisely pick the early Universe annihilation cross section necessary for the correct relic density but do not significantly affect the direct detection cross section). In the limit where $m_{\psi}$ is much larger than the mass of a xenon nucleus, direct detection constraints on the cross section scale with $m_{\psi}$, so our constraints on $g'\epsilon$ will scale like $m_{Z'}^2 \sqrt{m_{\psi}}$, which goes like $m_{Z'}^{5/2}$ in Fig.~\ref{fig:Direct Detection}.  

Complementary collider constraints exist from CMS and ATLAS searches for narrow dilepton resonances, 
see Fig.\,\ref{fig:Direct Detection}. The orange region labelled CMS is the constraint from searches for muon pairs in the 13 TeV data corresponding to an integrated luminosity of $137$ fb$^{-1}$ \cite{Sirunyan:2019wqq}.   This search provides direct bounds on $g^{\prime} \epsilon$ for $110$ GeV $< m_{Z^{\prime}} < 200$ GeV. 
The CMS bound at higher masses \cite{CMS:2019tbu} and ATLAS bound \cite{Aad:2019fac} require a conversion from bounds on a fiducial  cross section $\sigma_{fid} \times BR (Z' \to ll)$ reported in \cite{CMS:2019tbu} and \cite{Aad:2019fac}, respectively, to bounds on $g^{\prime} \epsilon$.  We do this by implementing the model using FeynRules \citep{Alloul_2014} and simulating via matching MadGraph5 \cite{Alwall_2014} with Pythia6 \cite{Sjorstrand_2006}. 
To extract these limits, we assume that no decays to the visible superpartners or hidden sector states are kinematically accessible, and that the $Z^{\prime}$ is narrow.  For a general branching ratio $BR_{SM}$ to SM states, the bound would be modified as $\epsilon \to \epsilon \ BR_{SM}^{-1/2}$, assuming the width remains modest.

\section{Indirect Detection}\label{sec:indirect}
We now comment on implications for indirect detection signals. The relic abundance is essentially determined via a “WIMP miracle,” and annihilations into hidden sector states are unsuppressed by $\epsilon$ and typically dominated by $s$-wave processes.   Thus, the present day dark matter annihilation cross sections can be large enough to make indirect detection a potentially powerful probe.   Hidden sector decays are rapid enough to be considered prompt for indirect detection signals.\,\footnote{Scenarios where this is not the case can lead to interesting signatures, see e.g.\cite{Rothstein:2009pm,Chu:2017vao,Kim:2017qaw,Gori:2018lem}.}

The indirect detection signals will be sensitive to the mass spectra in both hidden and visible sectors, as well as to whether R-parity is conserved or broken. The possibilities are numerous.  For now, we limit the discussion to qualitative comments, and leave a detailed treatment of various possibilities, including calculations of the precise spectra of SM final states and limits/projections from various experiments such as Fermi and CTA \cite{Doro:2012xx}, to future work \cite{toappear}. 

For fermionic dark matter,  the dominant annihilation channel over the vast majority of the parameter space is $\psi \bar{\psi} \rightarrow Z^{\prime} H^{\prime}$, followed by decay via portals to the visible sector as discussed in Section \ref{sec:decays}.  Our expectation is that the $H^{\prime}$ is sufficiently heavy that on-shell decays to $W$ bosons are accessible, whereas the $Z^{\prime}$ will decay to a mix of light SM fermions, typically up-type quarks and leptons.  
We therefore expect the dominant contribution to the photon spectrum from, \textit{e.\,g.}, the galactic center and dwarf galaxies to come from the hadronization of the quarks to pions and their subsequent decay; this is consistent with earlier works that studied indirect detection spectra of similar hidden sector cascade decays \cite{Elor:2015tva,Elor:2015bho,Escudero:2017yia}.\footnote{The work of \cite{Escudero:2017yia} attempted to fit a similar model consisting of $\sim$ 20 GeV dark matter cascading via ${\mathcal O}$(GeV) hidden bosons to the galactic center excess.}   

Modifications to this base case can occur when annihilations or decays to neutralinos become important. As discussed in Section\,\ref{sec:decays} (see also Fig.~\ref{fig:HiggsContour}), the $H^{\prime}$ will dominantly decay to $\chi_1^{\prime} \chi_1^{\prime}$ if kinematically allowed, or to $\chi_1\chi_1'$ in some regions of parameter space.  Alternatively, as shown in Fig.~\ref{fig:PsiAnnihilation}, the correct thermal relic abundance can be achieved by dark matter annihilations directly to $\chi_1'$. In such cases, we need to understand the fate of  the $\chi_{1}^\prime$, which decays via  portal couplings to $\chi_1$. Decays of the type $\chi_{1}^\prime \rightarrow \chi_1 V$  with $V= Z$, $h$ or $\chi_{1}^\prime \rightarrow \chi_1  f \bar{f}$ can dominate.  Thus, dark matter annihilations can take the form $\psi \bar{\psi}  \rightarrow \chi^{\prime} \chi^{\prime} \rightarrow \chi_1 \chi_1$  + $VV/4f$ or 
 $\psi \bar{\psi}  \rightarrow Z^{\prime} h^{\prime} \rightarrow (f\bar{f}) +  \chi^{(\prime)} \chi^{\prime} \rightarrow \chi_1 \chi_1$  + $V(V)/ 2(4)f$.  Further decays of $\chi_1$ into three SM fermions via the RPV coupling adds another step in the cascade.  Therefore, a single dark matter annihilation process could produce as many as 10 fermions in multiple steps. 

For scalar dark matter, in the case where one of the scalar masses is negative, $S_1 S_1 \rightarrow H^{\prime} H^{\prime}$ often dominates, with the $H^{\prime}$ decaying as discussed in Section \ref{sec:decays}. In the case where both scalar masses are positive and the hidden sector spectrum is more compressed, the dominant annihilation channel over much of the parameter space is $S_1 S_1 \rightarrow Z^{\prime} Z^{\prime}$, with the $Z^{\prime}$ primarily decaying into SM fermions.  This case also admits regions of parameter space where annihilation to neutralinos (and their attendant cascades, as described above) can be important, again leading to multiple SM fermions in the final state.  

The realistic hidden sector dark matter scenarios considered in this paper can therefore lead to more complicated signatures compared to ``simplified" hidden sector dark matter scenarios (such as those considered in \cite{Bell:2016fqf,Bell:2016uhg}), which generally consist of two-step dark matter annihilations of the form $DM+DM\to Z' H'\to 4f$.

\section{Conclusions}\label{sec:conclusion}

In this paper, we have put together a simple framework for dark matter, building upon ingredients and guiding principles that are  well-motivated: hidden sectors, supersymmetry, naturalness, and the realization of the correct relic density for dark matter via the WIMP miracle.  A hidden sector can lie around the weak scale, thereby realizing the WIMP miracle.  This happens naturally in scenarios where, for instance, supersymmetry breaking is mediated to both the hidden and visible sectors via gravity mediation. This can be made compatible with stringent LHC limits on superpartners with only $\mathcal{O}(1)$ differences between hidden and visible sector parameters in the UV. In this framework, we studied the minimal matter field content under a hidden sector $U(1)'$ gauge symmetry that kinetically mixes with the SM hypercharge, where dark matter is stabilized not by R-parity but by an accidental $Z_2$ symmetry in the hidden sector. While the electroweak scale in our sector might be accidentally small, we assumed symmetry breaking in the hidden sector to be ``natural," which suggests that the hidden sector scalars and fermions as well as their superpartners lie around the same mass scale, opening possibilities for a variety of dark matter candidates as well as rich cosmological histories and indirect detection signatures. 

For fermion dark matter, we found that dark matter annihilation is generally dominated by the $Z'H'$ channel, though annihilations to hidden sectors neutralinos are still relevant. 
For scalar dark matter, annihilations to $\chi_i' \chi_j'$, $H'H'$ as well as $Z'Z'$ states were shown to lead to consistent cosmological histories. We also found instances of coannihilation between the scalar and the fermion providing the correct relic density, where the heavier of the two can be extremely long-lived, well beyond BBN, yet consistent with all cosmological constraints. 

In such frameworks, dark matter direct detection cross sections and production cross sections for hidden sector particles at colliders are generally suppressed by the portal coupling strength $\epsilon$ mixing the two sectors. While such signals might be observed, a too-small $\epsilon$ would preclude such possibilities.  Indirect detection is different: dark matter annihilation into visible particles proceeds via a series of cascade decays involving hidden sector particles, and can lead to a wide variety of indirect detection signals that might be within reach of future experiments; detailed studies of such signals will be performed in a future paper \cite{toappear}.

\section*{Acknowledgements}
We thank Stefania Gori for helpful conversations. The work of PB, ZJ and AP was supported by the DoE under grant DE-SC0007859.   BS thanks the GGI Institute for Theoretical Physics, the Mainz Institute for Theoretical Physics (MITP) of the Cluster of Excellence PRISMA+ (Project ID 39083149), and the Leinweber Center for Theoretical Physics (LCTP) at the University of Michigan, where parts of this research were completed, for hospitality and support.

\bibliography{HiddenWIMP} 

\section{Appendix}
\appendix 

We first  present analytic one-loop solutions to the RGEs of the model considered in this paper.  We define $t \equiv log(\mu/M_{GUT})$, with $M_{GUT} \simeq 2 \times 10^{16}$ GeV.
  UV boundary conditions will be specified with a $0$ subscript at $t=0$.     
  \begin{align}
\a'(t)&=\frac{\a'_0}{1-\a'_0 t/\pi}\,,\\
\a_\l(t)&=\alpha_{\l_0} \frac{-4 \pi F'(t)}{1+6\alpha_{\l_0} F(t) }\,,\\
F(t)&= -\frac{t}{12\pi}\left(3 - 3 \frac{\a'_0 t}{\pi} + \frac{\a_0^{\prime 2} t^2}{\pi^2}\right).\label{eqn:Ft}
\end{align}
Note that since $t<0$ in the IR, $F(t)>0$ and increases monotonically.
\begin{figure}[t]
	\centering
	\begin{minipage}{.49\textwidth}
	\includegraphics[width=1\columnwidth]{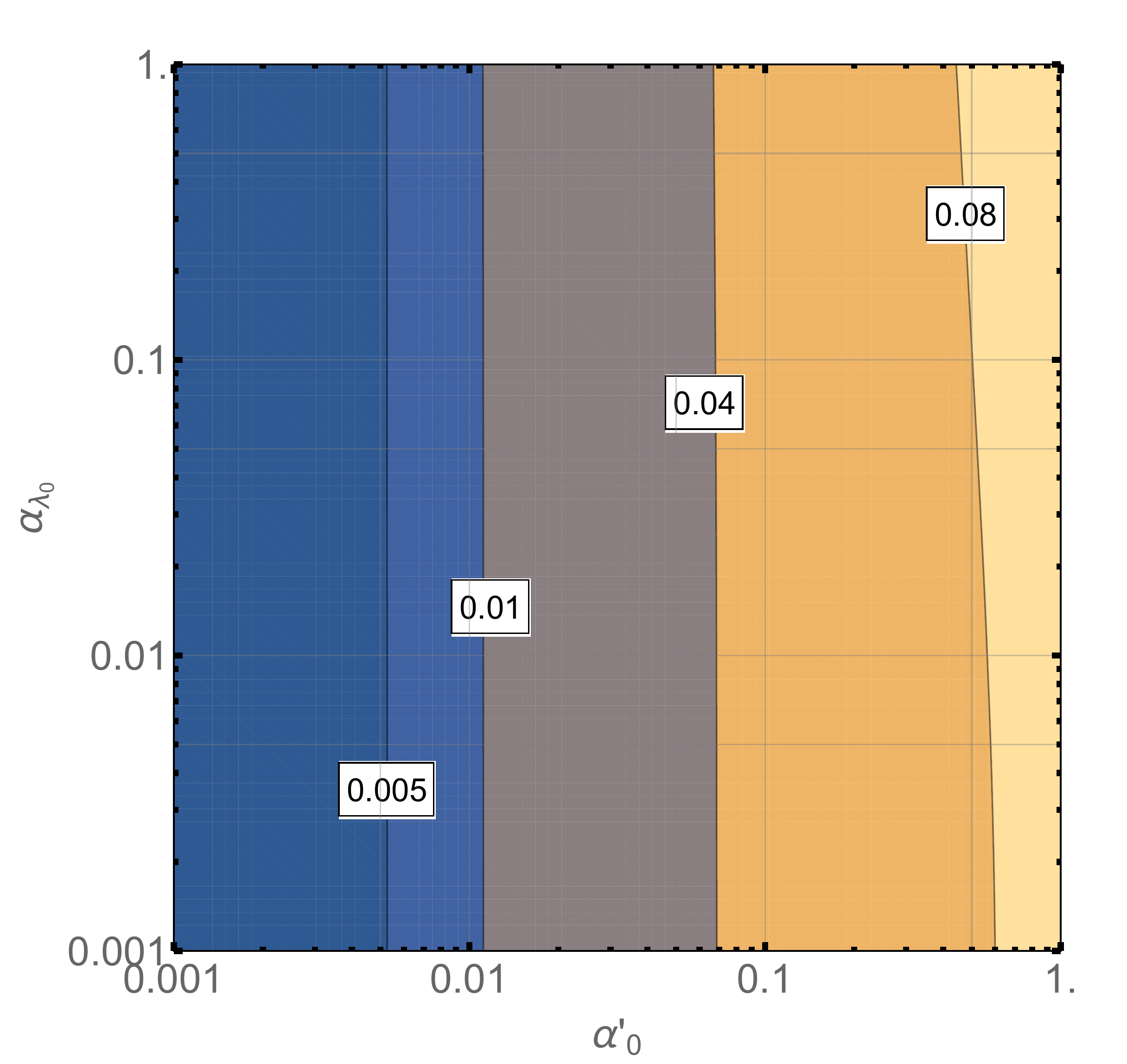}\\\textbf{a) $\a^{\prime \text{IR}} $}
	\end{minipage}%
	~~
	\begin{minipage}{.49\textwidth}
	\includegraphics[width=1\columnwidth]{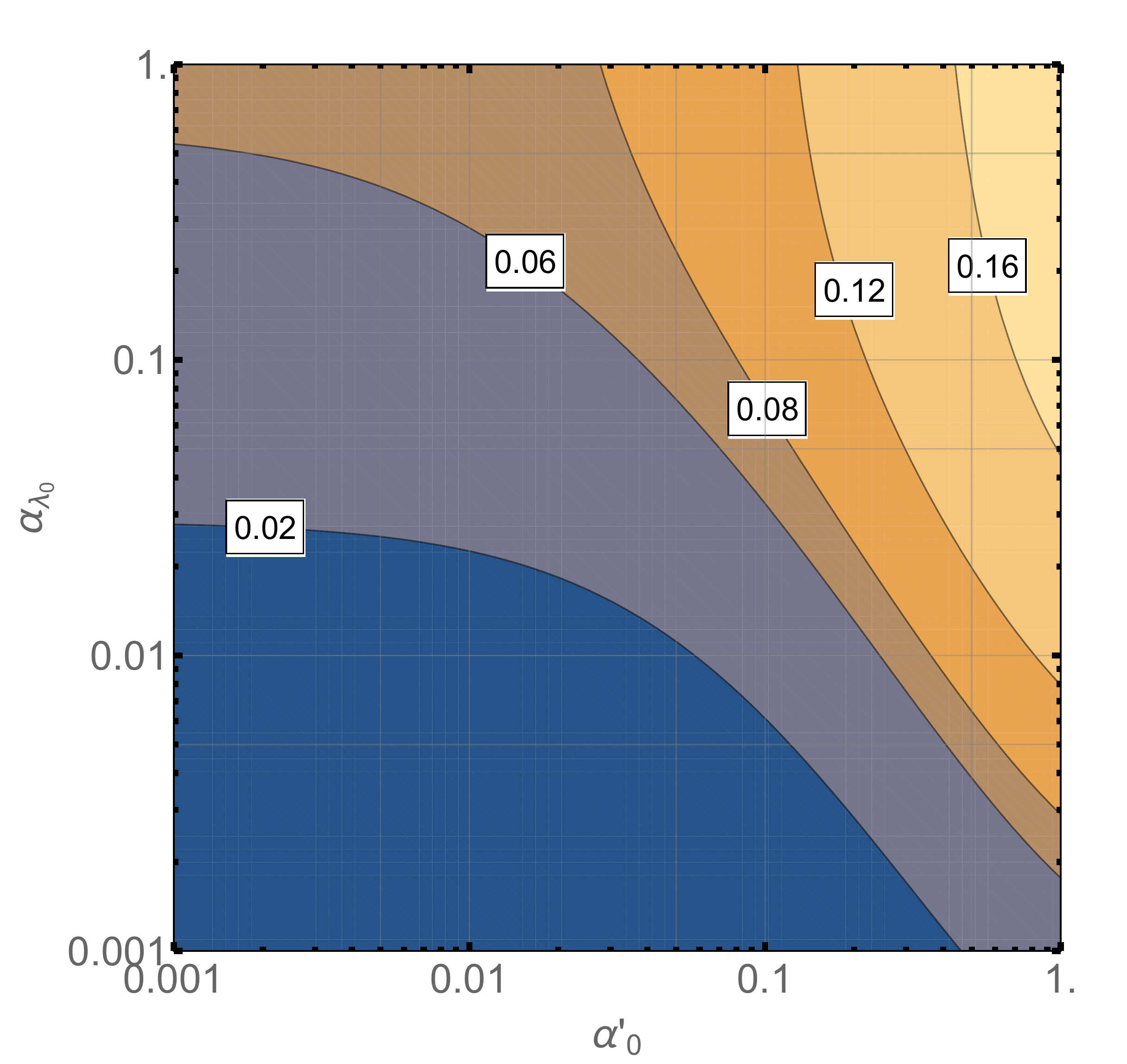}\\\textbf{b) $\a_\l^{\text{IR}} $}
	\end{minipage}
	
	\caption{The IR couplings resulting from a two-loop numerical RG flow from $M_{GUT}$ to $m_{top}$. \textbf{Left Panel:} The gauge coupling $\alpha'$. \textbf{Right Panel:} The Yukawa coupling $\a_\l$.}
	\label{fig: IRCouplings}
\end{figure}
In this Appendix, when we display quantities in the IR, for concreteness, we evaluate them at $\mu=m_{top}$.

\begin{figure}
	\centering\  
	\textbf{ $r^{1/2}=\sqrt{(1-m_{Z'}^2/m_\psi^2)} $
	\centering\ \\\
	\includegraphics[width=0.5\columnwidth]{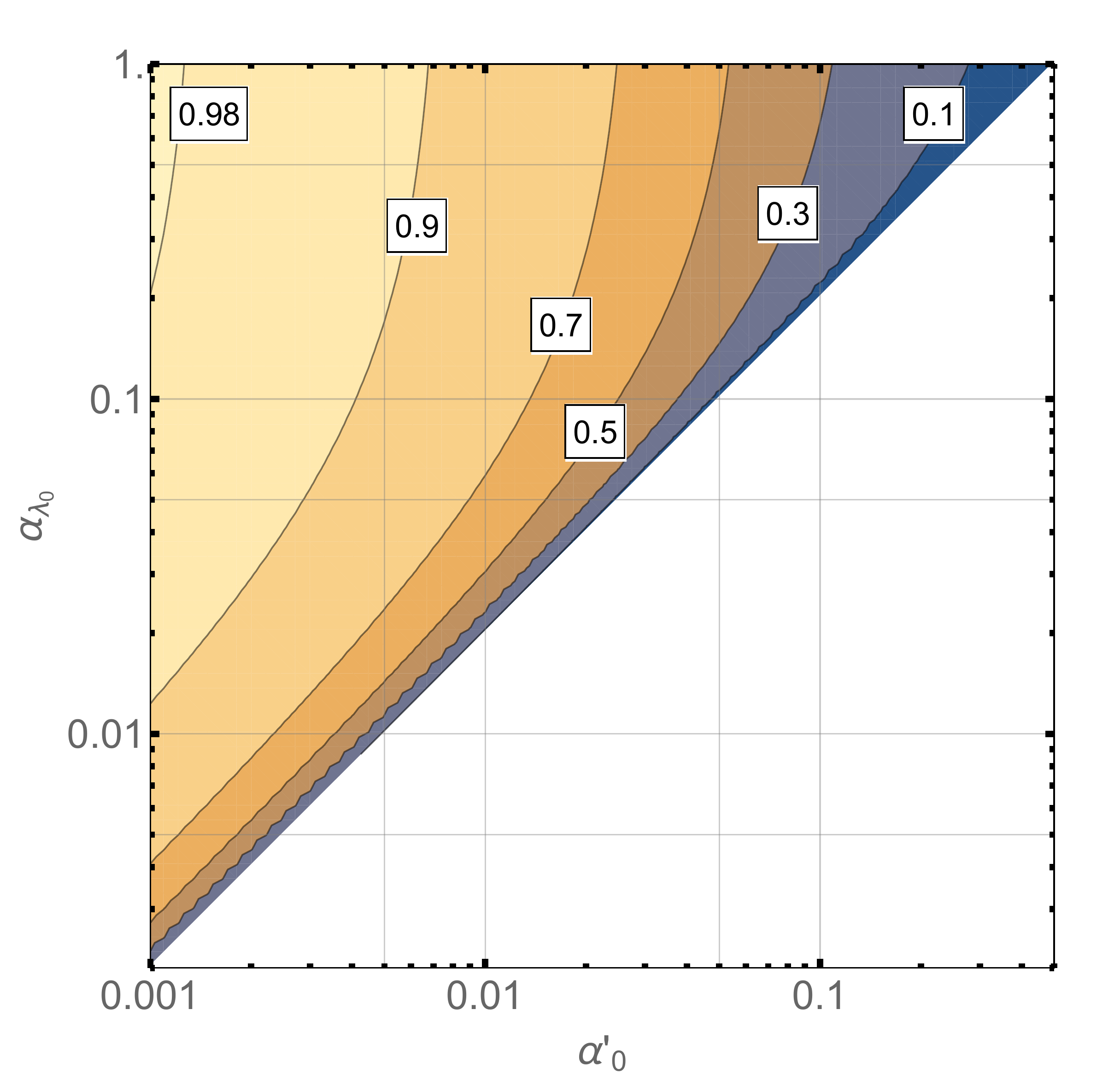}}
	\caption{Contours of the IR kinematic suppression of $\psi$ annihilation to $m_{Z'}$ as a function of UV couplings. This suppression is exactly $r^{1/2}$, for $r$ defined in Eq.\,(\ref{eqn:rEvolution}).}
	\label{fig:kinematic_Suppression}
\end{figure}

The two-loop numerical RG flow is shown in Fig. \ref{fig: IRCouplings}, and is well approximated by the above formulae.  Recall that these couplings have a fixed point at $2\a'=\a_\l $, where the $\psi$ becomes degenerate with the $Z',\,H'$.  We display the impact of this fixed point on the possible kinematic suppression of fermion annihilation to $Z^{\prime} H^{\prime}$ in Fig.~\ref{fig:kinematic_Suppression} (as discussed near Eq.~(\ref{eqn:rEvolution})\,).  For large couplings, it is possible that evolution towards the fixed point can cause substantial suppression of the annihilation rate into the $Z^{\prime} H^{\prime}$ final state.  

The coupling evolution itself is enough to determine two more RG parameters:
\begin{align}
m_{\tilde{B}'}(t)&=m_{\tilde{B}'_0} \frac{\a'(t)}{\a'_0}\,,\\
A_\l(t)&=A_{\l_0}\frac{1}{1+6\a_{\l_0} F(t) } + m_{\tilde{B}'_0} \left( \frac{2t\a'(t)}{\pi} + 6 \a_{\l_0} \frac{t F'(t)-F(t)}{1+6\a_{\l_0} F(t) }   \right).
\end{align}

In the basis 
\begin{eqnarray}
\Delta_{HT} &\equiv& \tilde{m}_{H'}^2 - \tilde{m}_T^2\,, \\
\Delta_{STH} &\equiv& 2 \tilde{m}_S^2 - (\tilde{m}_T^2 + \tilde{m}_{H'}^2)\,,\\
\Sigma &\equiv&  \frac{1}{3}(\tilde{m}_S^2 + \tilde{m}_T^2 + \tilde{m}_{H'}^2),
\end{eqnarray}
the solutions for the soft scalar masses can be captured by relatively simple expressions 
\begin{eqnarray}
\label{eqn:basissol}
\Delta_{HT}(t)&=&\Delta_{HT_0} \frac{\a'(t)}{\a'_0}\,,\\
\Delta_{STH}(t)&=&\Delta_{STH_0} - 2 m^2_{\tilde{B}'_0} \left(1-\frac{\a'(t)^{2}}{\a_0^{\prime 2}}\right)\,,\\
\Sigma&=&\Sigma_{STH_0}\frac{1}{1+6\a_{\l_0} F(t) } + \frac{2}{3} m^2_{\tilde{B}'_0} \left(1-\frac{\a'(t)^{2}}{\a_0^{\prime 2}} \right)- 2 I(t),
\end{eqnarray}
where we have defined $I(t)$ as 
\begin{eqnarray}
I(t)&=& m_{\tilde{B}'_0} ^2 I_{GG}(t) + m_{\tilde{B}'_0}  A_{\l_0} I_{AG}(t) + A_{\l_0}^2 I_{A A}(t)\,,
\end{eqnarray}
where
\begin{eqnarray}
I_{AA}(t)&=&\frac{\alpha_{\lambda_0} F(t)}{(1+6\alpha_{\lambda_0} F(t))^2}\,,\\
I_{AG}(t)&=&-2\frac{\alpha_{\lambda_0} (t F'(t)-F(t))}{(1+6\alpha_{\lambda_0} F(t))^2}\,,\\
I_{GG}(t)&=&-6\frac{\alpha_{\lambda_0}^2 (t F'(t)-F(t))^2}{(1+6\alpha_{\lambda_0} F(t))^2} + 
\frac{\alpha_{\lambda_0} G(t)}{1+6\alpha_{\lambda_0} F(t)}\,,\\
G(t)&=& -\frac{2}{\pi} \a'(t) t^2 F'(t),
\end{eqnarray}
with $F(t)$ defined in Eq.~(\ref{eqn:Ft}).
The above equations can be inverted to yield solutions for the soft masses.

We have verified these analytic results against numerical solutions of the full two-loop RGE.   The two-loop RGEs have previously been discussed in \cite{Andreas:2011in}; we correct a few small typographical errors: in Equation (A1) of Ref.~\cite{Andreas:2011in}, in the two-loop part of the $\beta$ function for the gaugino mass there is an  $\alpha_{h}^2$ that should read $\alpha_{h}$; in the two-loop part of the $\beta$ function for $m_{S}$ there is an $\alpha_S^2$ that should read $\alpha_S$, and in the two loop expression for the $\beta$ function for $m_{\pm}$ there is an $\alpha$ that should read $\alpha_{h}$.

\begin{figure}
\includegraphics[width=1\columnwidth]{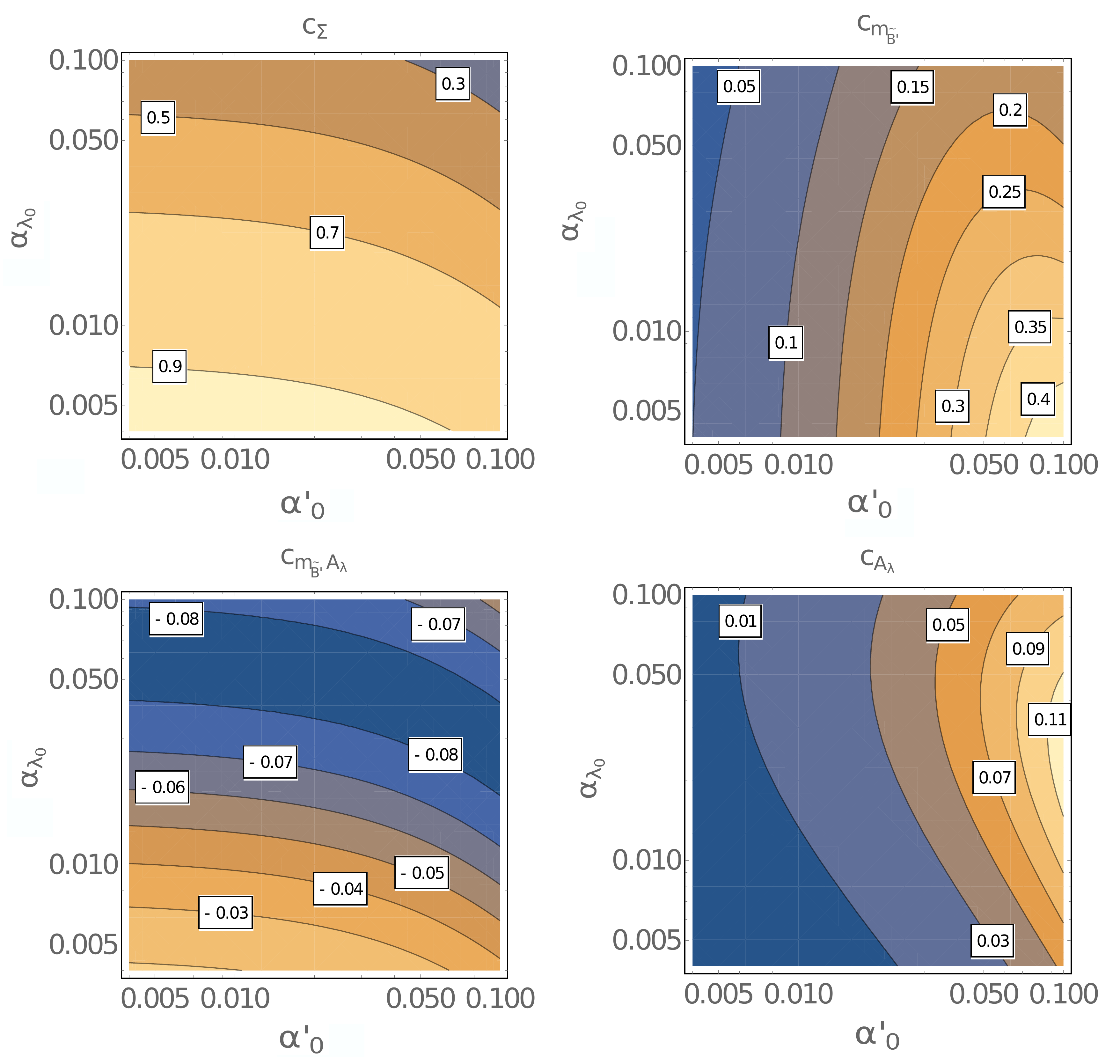}
\caption{Each plot depicts the IR coefficient weighting respective soft breaking terms whose linear combination gives the IR value of $\Sigma$, defined in Eq.~(\ref{eq:appendix-Sigma}). The axes show the UV couplings that determine the coefficients (up to $\epsilon$-suppressed effects). }
\label{fig:IR_Sigma-scale}
\end{figure}

At one loop, our proxy for the hidden sector scale,  $\Sigma$, when evaluated in the IR, can be written in terms of UV parameters as:
\begin{equation}
\Sigma = c_\Sigma \Sigma_0 + c_{m_{\tilde{B}'}}m^2_{\tilde{B}'_0}  + c_{m_{\tilde{B}'} A_\l} m_{\tilde{B}'_0} A_{\l_0} + c_{A_\l} A^2_{\l_0}.\label{eq:appendix-Sigma}
\end{equation}
In Fig.~\ref{fig:IR_Sigma-scale}, we display these coefficients as functions of the dimensionless parameters $\alpha^{\prime}$ and $\alpha$ in the UV.  We note that $c_{m_{\tilde{B}'}}$ and $c_{\Sigma}$ are the largest numerically, thus we  expect UV specification of the gaugino mass and/or $\Sigma$ will largely determine the scale of the hidden sector in the IR, absent very large $A$-terms in the UV.  Furthermore, examination of the values of the $c_{m_{\tilde{B}'}}$ and $c_{\Sigma}$ in the figure, when taken in concert with analogous expressions for the MSSM, see Eq.~(\ref{eqn:MSSMUVIR}), shows the disparity between the importance of the UV gaugino mass in these two sectors; the MSSM is much more sensitive to UV gaugino masses due to the strongly coupled $SU(3)$ in the IR.

It is interesting to note the existence of additional IR fixed points in this model. The parameters $m_{\tilde{B}'}$ and $\Delta_{HT}$ defined above flow to zero in the IR with identical one-loop solutions. However, this is a relatively slow effect, as can be extracted from Eq.~(\ref{eqn:basissol}) and Fig.\ref{fig: IRCouplings}. The trilinear $A_\l$ term also flows to a fixed point given by $A_\l=-\frac{2}{3} \frac{\alpha}{2 \alpha_\lambda} m_{\tilde{B}'}= -\frac{2}{3} m_{\tilde{B}'} $.

\end{document}